\documentstyle[11pt,epsfig]{article}
\def\la{\langle}\def\ra{\rangle}
\def\be{\begin{eqnarray}}\def\bea{\begin{eqnarray}}
\def\ba{\begin{eqnarray}}
\def\ee{\end{eqnarray}}\def\eea{\end{eqnarray}}
\def\ea{\end{eqnarray}}
\def\ben{\begin{eqnarray}}\def\bitem{\begin{itemize}}
\def\een{\end{eqnarray}}\def\eitem{\end{itemize}}
\def\del{\partial}
\def\G0p{$G_0^\prime$}
\def\bi{\bibitem}

\def\N1520{$N^\star (1520)$}
\def\calL{{\cal L}}

\def\prl{Phys. Rev. Lett.}\def\pr{Phys. Rev.}\def\np{Nucl. Phys.}
\def\pl{Phys. Lett.}
\def\B#1{{}^{#1}\mbox{B}}

\def\del{\partial}

\def\roughly#1{\mathrel{\raise.3ex\hbox{$#1$\kern-.75em%
\lower1ex\hbox{$\sim$}}}}\def\lsim{\roughly<}
\def\gsim{\roughly>}

\def\B#1{{}^{#1}\mbox{B}}

\def\itt{\indent\indent}

\def\B0{\mbox{\boldmath $0$}}

\def\bq{\begin{equation}}
\def\eq{\end{equation}}

\def\b3{\mbox{\boldmath $3$}}\def\b6{\mbox{\boldmath $6$}}
\def\tr{\rm tr}
\def\rhosobar{$\rho$-sobar}\def\nstar{$N^* (1520)$}

\renewcommand{\thefootnote}{\fnsymbol{footnote}}
\begin{document}
\begin{titlepage}
\begin{center}

{\LARGE \bf  Double Decimation and Sliding Vacua\\ in the Nuclear
Many-Body System}
  \vskip 2cm
   {{\large G.E. Brown$^{(a)}$ and Mannque Rho$^{(b,c)}$} }
 \vskip 1cm

{(a) \it Department of Physics and Astronomy, State University of
New York,\\ Stony Brook, NY 11794, USA}

 {(b) \it  Service de Physique Th\'eorique, CEA/DSM/SPhT,
Unit\'e de recherche associ\'ee au CNRS, CEA/Saclay,  91191
Gif-sur-Yvette c\'edex, France}

{(c) \it School of Physics, Korea Institute for Advanced Study,
Seoul 130-722, Korea}

\end{center}

\vskip 1cm

\centerline{(\today)}
 \vskip 1cm

\centerline{\bf Abstract}
 \vskip 0.5cm
We propose that effective field theories for nuclei and nuclear
matter comprise of ``double decimation": (1) the chiral symmetry
decimation (CSD) and (2) Fermi liquid decimation (FLD). The
Brown-Rho scaling identified as the $parametric$ dependence
intrinsic in the ``vector manifestation" of hidden local symmetry
theory of Harada and Yamawaki results from the first decimation,
i.e., CSD. In a recent work we showed that in matter under
conditions of high density or high temperature, dynamically
generated hadron masses scaled with a common scale. Namely for low
densities and temperatures the masses scaled as $m^\star/m\simeq
[\la\bar{q}q\ra^\star/\la\bar{q}q\ra]^{1/2}$ whereas at higher
densities and temperatures as
$[\la\bar{q}q\ra^\star/\la\bar{q}q\ra]$. In the present work we
summarize new empirical evidence for Brown-Rho (BR) scaling and
discuss in a general way its impact on the nuclear many-body
problem. While the double decimation should be carried out, it has
been a prevalent practice in nuclear physics community to proceed
with the second decimation, assuming density independent masses,
without implementing the first, chiral symmetry decimation. We
show why most nuclear phenomena can be reproduced by theories
using either density-independent, or density-dependent masses, a
grand conspiracy of nature that is an aspect that could be tied to
the Cheshire-Cat phenomenon in hadron physics. We go through some
of the history of the chiral symmetry decimation (CSD) which
involves BR scaling, in order to show that historically one had to
look at very specific phenomena involving transition matrix
elements to see that it is necessary. We identify what is left out
in the Fermi liquid decimation (FLD) that does not incorporate the
CSD. Experiments such as the dilepton production in relativistic
heavy ion reactions, which are specifically designed to observe
effects of dropping masses, could exhibit large effects from the
reduced masses. However they are compounded with effects that are
not directly tied to chiral symmetry. We discuss a recent
STAR/RHIC observation where BR scaling can be singled out in a
pristine environment.

\end{titlepage}
\newpage
\renewcommand{\thefootnote}{\arabic{footnote}}
\setcounter{footnote}{0}
\section{Introduction}\label{intro}
\itt Brown and Rho adduced empirical evidence for BR scaling in
terms of nuclear interactions and nuclear structure more than a
decade ago~\cite{BR-e89,BR-stiff90,BR91}. One of the most
convincing evidences was the decrease in the tensor interaction in
nuclei: The $\pi$ and $\rho$ exchange enter into the tensor
interaction with opposite signs. The $\rho$-coupling to the
nucleon is about twice that of vector dominance, so its square is
four times greater. Brown and Machleidt~\cite{brown-machleidt}
showed this to follow unambiguously from the phase shifts from
nucleon-nucleon scattering.

The large $\rho$ tensor coupling is important in free space, in
that it cancels the otherwise extremely strong pionic coupling,
which goes as $r^{-3}$ at short distances, at a distance of $\sim
0.6$ fm. Inside of this, the net tensor interaction is repulsive,
but strongly cut down in net effect by the large repulsion from
$\omega$-exchange which keeps the two nucleons apart. Because of
the $\rho$ exchange, the nucleon-nucleon interaction never gets
very strong, explaining why the effective nucleon-nucleon
interaction obtained in the Sussex work~\cite{sussex}, which
neglected off-shell effects, worked well in reproducing the
effective nucleon-nucleon interaction in nuclei. The very smooth
behavior of the nuclear interactions in going off shell minimizes
the difference between the on-shell interaction and the half
off-shell one that is commonly calculated as effective
interactions.

{\it In medium}, with the decrease in $m_\rho^\star$ to $\sim 0.8
m_\rho$ by nuclear matter density $n_0$, the tensor interaction
becomes even weaker, and this was the early evidence adduced by
Brown and Rho for BR scaling.

Our work is prompted by two recent developments that are both
profound and powerful for nuclear structure. One is the notion of
``vector-manifestation fixed point" in effective field theory of
hadrons discovered by Harada and Yamawaki~\cite{HY:VM,HY:PR} and
the other is the identification of nuclear matter as a Fermi
liquid at its fixed point~\cite{shankar,friman,songetal}. The
first fixed point is dictated by the matching of effective field
theories to QCD and the second accounts for the stability of
strongly interacting many-body systems that possess a Fermi
surface arising from a quantum critical phenomenon. We discuss how
one could combine these two fixed point structures into an
effective field theory of nuclei and nuclear matter.

The plan of this review is as follows.

In Section \ref{doubledecimation} we develop the concept of
``double decimation" for describing nuclear phenomena based on the
general strategy of effective field theories that can represent
low-energy nonperturbative QCD. The first decimation deals with
chiral symmetry scale and we suggest that the ``parametric"
density dependence encoded in BR scaling~\cite{BR91} is the
consequence of the ``chiral scale decimation" (CSD) tied to the
``vector manifestation" \`a la Harada and
Yamawaki~\cite{HY:VM,HY:PR} in hidden local symmetry theory that
is matched to QCD. The second decimation deals with the Fermi
momentum scale and corresponds to (Landau) Fermi liquid
description of nuclear matter. We shall refer to this as ``Fermi
liquid decimation (FLD)."

 In Section \ref{eft}, one way of doing effective
field theory in nuclear physics is presented with a focus on how
to incorporate the standard nuclear physics approach (SNPA) that
has had a spectacular success, into the framework of modern
effective field theory that we shall refer to as {\it more
effective} effective field theory (MEEFT in short). Both the
effective nucleon interaction $V_{low-k}$ and the electroweak
matrix elements of few-nucleon systems are described in this
formalism.

In Section \ref{diracpheno} we discuss in a general way why the
need for density dependent hadron masses was missed for so many
years in nuclear phenomena; namely, the description of most
phenomena with density independent masses could fit experiments.
We show that with a common scaling with density of hadron masses,
the kinematics of the nuclear many-body problem are not changed
very much in the region of densities up to nuclear matter density
and we illustrate this with the highly successful Dirac
phenomenology.

As cases where BR scaling can be singled out in a specific way in
nuclear processes, we discuss in Section \ref{axialcharge} how the
scaling enters in what is known as Warburton's $\epsilon_{MEC}$
factor in axial charge transitions in heavy nuclei as well as in
deeply bound pionic atoms which have been recently studied
experimentally.

In Section \ref{photon}, we suggest that the vector manifestation
fixed point $a=1$ of Harada and Yamawaki is ``precociously"
reached when baryons are present and hence in nuclear medium, as a
consequence of which the photon couples half and half to the
vector meson and the nucleon core (or skyrmion). This picture may
be relevant to the electromagnetic form factors of the proton
recently measured at the J-Lab as well as to the longitudinal and
transverse form factors of nuclei with the meson and nucleon
masses dropping \`a la BR scaling.

While it is now generally accepted that the CERES dileptons can be
explained by modified properties of hadrons in dense and hot
matter, it is difficult to single out BR scaling as the principal
mechanism for the shifted spectral distribution as temperature and
density effects are compounded in the process. We sketch in
Section \ref{sobar} what we believe to be the correct way of
interpreting the dilepton data.

It is discussed in Section \ref{RHIC} how the situation can be a
lot clearer in the RHIC experiments. We discuss in particular how
a direct measurement of medium-dependent vector-meson mass $m_V^*$
could be made in a pristine environment where temperature effects
are small and where the density can be well reconstructed.

A discussion of the (absence of) need for ``double decimation" in
nuclear structure makes up Section \ref{effectiveforce} where we
discuss effective interaction between nucleons in nuclei. The
Kuo-Brown interactions~\cite{kuo-brown} have undergone a
renaissance with the finding from the renormalization-group (RG)
formulation that all nucleon-nucleon interactions which fit the
two-body phase shifts give the same effective interactions
$V_{low-k}$ in nuclei. The RG cutoff $\Lambda$ is chosen to be at
the upper scale to which phase shifts have been obtained from
nucleon-nucleon scattering; i.e. $\Lambda\sim 350$ MeV.

That $V_{low-k}$ provides a unique effective interaction which
works well in reproducing effective forces in nuclei is
unquestionable, given the excellent fits to all sorts of data.
Nonetheless, there are questions:

\noindent$\bullet$ {\bf (1)} Where does the modification in the
tensor interaction, which is lowered by the decreasing mass of the
$\rho$-meson which contributes with opposite sign to the pion
(whose mass is presumably unchanged in medium at low density),
come in? It is well known that certain states, such as the ground
state of $^{14}$N, depend sensitively on the tensor interaction.
Also in the nuclear many body problem, the decrease in the tensor
force will give less binding energy and less help with saturation,
the second order tensor force giving a large contribution to the
binding energy of nuclear matter. In fact, this second order term
which has the form
 \be
V_{eff}\simeq
\left(\frac{3+\vec{\sigma}_1\cdot\vec{\sigma}_2}{4}\right) V(r)
 \ee
is large as we discuss. In an old paper, Feshbach and
Schwinger~\cite{feshbach} describe the difference between $^1S$
and $^3S$ potentials in terms of the above $V_{eff}$ which acts
only in the latter state.

\noindent$\bullet$ {\bf (2)} In addition to the modifications in
both tensor and spin-orbit interactions, which have not been put
into the Fermi liquid decimation, there is the question of
off-shell energies of intermediate particles~\footnote{The holes
are on-shell in the Brueckner-Bethe theory.} which are taken to be
plane waves. The conclusion following from the Bethe reference
spectrum~\cite{gebbook} was that there was a small amount of
binding in the intermediate states with energies just above the
Fermi surface, referred to as ``dispersion" correction. Generally
taking the intermediate states to be free, as in the Schwinger
interaction representation, seemed to be a good approximation.
This is what is done in the RG calculations, in the Fermi liquid
decimation.

Concluding remarks are given in Section \ref{concl} with a
reference to the Cheshire Cat Principle revisited. We postulate
that the missing ``smoking gun" for BR scaling in nuclear
structure physics is an aspect of the Cheshire Cat Principle
discovered in the baryon structure as reflected in many-nucleon
systems.

\section{The Double Decimation}\label{doubledecimation}
\itt Our ultimate objective is to describe in a unified way finite
nuclei, nuclear matter and dense matter up to chiral restoration.
For this we introduce the approximation -- double decimation -- by
which the phase structure in the hadronic sector can be
drastically simplified. The effective field theory (EFT) approach
to few-nucleon systems described below does not require both
decimations but if one wants to correctly describe phase
transitions of hadronic systems under extreme conditions of
density and/or temperature, one needs at least two decimations
which we now describe. We propose that the procedure consists of
what we call ``chiral symmetry decimation (CSD)" and ``Fermi
liquid decimation (FLD)."
\subsection{The ``intrinsic dependence (ID)"
from chiral symmetry decimation}\label{parametric}
 \itt As we have explained in a series of recent
papers~\cite{BR:PR01,BR:BERK,MR:taiwan,BR:qm02}, a candidate
effective field theory relevant in the hadronic sector matched to
QCD at a scale near the chiral scale $\Lambda_\chi$ is hidden
local symmetry (HLS) theory with the vector manifestation (VM)
found by Harada and Yamawaki~\cite{HY:VM,HY:PR}. This theory,
denoted HLS/VM in short, with the vector mesons~\footnote{We are
assuming $U(2)$ flavor symmetry in the two-flavor case.}, $\rho$,
$\omega$ etc. treated as $light$ on the same footing as the
(pseudo) Goldstone pions as originally suggested by
Georgi~\cite{georgi}, turns out to yield results consistent with
chiral perturbation theory~\cite{HY:WM}. Given this theory valid
at low energy in matter-free space, one can then construct the
chiral phase transition in both hot~\cite{HaradaSasaki} and
dense~\cite{HKR} media recovering BR scaling~\cite{BR91} near
chiral restoration. This theory makes unambiguous predictions, at
least at one-loop order, on the vector and axial-vector
susceptibilities and the pion velocity at the transition point
that seem to be different from the standard scenario and that can
ultimately be tested by QCD lattice measurements~\cite{hkrs}.

In the rest of this section we will confine out consideration on
finite density at zero temperature~\footnote{Implementing
temperature effects is straightforward and hence will be ignored
in what follows in this section.}.

An early, seminal attempt to arrive at nuclei and nuclear matter
in chiral Lagrangian field theory that models QCD was made by
Lynn~\cite{lynn}. What we are interested in here is to exploit the
recent development of HLS/VM and arrive at a field theoretic
description anchored on QCD. In studying nuclear systems, the EFT
and QCD have to be matched via current correlators at a suitable
scale $\Lambda_M$ in the background of matter characterized by
density $n$. The matching defines the ``bare" HLS Lagrangian,
giving to the parameters of the Lagrangian the ``intrinsic"
density dependence (IDD). This means that the parameters of the
Lagrangian such as hidden gauge coupling constant $g^*$, gauge
boson mass $m_\rho^*$ etc will depend intricately not only on
$\Lambda_M$ but also on density $n$. This intrinsic density
dependence is generally missing in the calculations that do not
make the matching to QCD. Next to account for quantum effects, we
need to decimate the degrees of freedom and excitations from the
matching scale $\Lambda_M$ to a scale commensurate with the
presence of a Fermi sea, $\Lambda_{FS}$. The latter is the scale
relevant to nuclear physics, typically $\Lambda_{FS}\lsim 1.5$
fm$^{-1}$. What governs this first decimation is chiral symmetry.
\subsubsection{Sliding vacua}\label{slidingvacua}
\itt  The HLS/VM at the present stage of development can tell us
what happens only at $n\approx 0$ and at $n\approx n_c$. It does
not tell us how the intrinsic density dependence (IDD)
interpolates from $n=0$ to $n=n_c$. What we are interested in,
however, involves a wide range of densities, from zero to nuclear
matter and ultimately to chiral restoration. How to do this is at
present poorly known. There are however two approximate ways to
address this issue.  One is based on extending NJL model so as to
simulate the vector manifestation effect associated with
vector-meson degrees of freedom. This is the approach we have
proposed before~\cite{BR:qm02} which we shall follow in the rest
of the paper. The other which we shall not employ in this paper is
however quite instructive on how the ``vacuum change" induced by
matter is manifested in many-nucleon systems. It is based on a
skyrmion description of dense hadronic matter which we briefly
describe.

The skyrmion approach to dense matter developed
recently~\cite{LPRVchi,LPMRV1,LPMRV2} makes transparent the
crucial role of the intrinsic density dependence (IDD) in nuclear
processes. The power of the skyrmion model is that a single
effective Lagrangian provides a unified description of mesons and
baryons and treats single-baryon and multi-baryon interactions on
the same footing, thereby describing finite nuclei as well as
infinite nuclear system -- nuclear matter. Although the notion of
skyrmions as modelling of QCD is fairly well
established~\cite{CND}, one does not yet know at present how to
write down a fully realistic skyrmion Lagrangian, and hence cannot
do a quantitative calculation but one can gain a valuable insight
into the structure of the ``sliding vacua" that we are interested
in.

Following \cite{LPRVchi,LPMRV1,LPMRV2}, one considers a
skyrmion-type Lagrangian with spontaneously broken chiral symmetry
and scale symmetry, associated, respectively, with nearly massless
quarks and trace anomaly of QCD. The effective fields involved are
the chiral field $U=e^{i\pi/f}$ with $\pi$ the (pseudo)-Goldstone
bosons for the former and the scalar ``dilaton" field $\chi$ for
the latter. Such a theory which may be considered to be an
$N_c\rightarrow \infty$ approximation to QCD can, albeit
approximately, describe not only the lowest-excitation, i.e.,
pionic, sector but also the baryonic sector and massive vector
meson sector all lying below the chiral scale
$\Lambda_\chi$~\footnote{This naturally implies that the gluonium
degrees of freedom lying higher than $\Lambda_\chi$ are to be
integrated out leaving only the ``soft" quarkonium degrees of
freedom. As shown in \cite{LPRVchi}, these are the degrees of
freedom intricately tied to the spontaneous breaking of chiral
symmetry.}. What is even more significant, it can treat on the
same footing both single-baryon and multi-baryon systems including
infinite nuclear matter~\cite{LPMRV1}.

We are interested in how low-energy degrees of freedom in
many-body systems behave in dense matter, in particular as the
density reaches a density at which QCD predicts a phase transition
from the broken chiral symmetry to the unbroken chiral symmetry,
i.e., chiral restoration. The simplest possible Lagrangian with
the given symmetry requirements consistent with QCD is of the
form~\cite{LPRVchi}
\begin{eqnarray}
{\cal L} &=& \frac{f_\pi^2}{4} \left(\frac{\chi}{f_\chi}\right)^2
{\rm Tr} (\partial_\mu U^\dagger \partial^\mu U) +\frac{1}{32e^2}
{\rm Tr} ([U^\dagger\partial_\mu U, U^\dagger\partial_\nu U])^2
+\frac{f_\pi^2m_\pi^2}{4} \left(\frac{\chi}{f_\chi}\right)^3 {\rm
Tr} (U+U^\dagger-2)
\nonumber \\
&& \hspace{2cm} +\frac{1}{2}\partial_\mu \chi\partial^\mu \chi
-\frac{1}{4}\frac{m_\chi^2}{f_\chi^2}\bigg[\chi^4\bigg(\ln(\chi/f_\chi)
-\frac{1}{4}\bigg)+\frac{1}{4}\bigg].
 \label{lag-chi}
\end{eqnarray}
We have denoted the vacuum expectation value of $\chi$ as
$f_\chi$, a constant which describes the decay of the scalar
$\chi$ into pions in matter-free space. The QCD trace anomaly can
be reproduced by the last term of (\ref{lag-chi}), i.e., the
potential energy $V(\chi)$ for the scalar field, which is adjusted
so that $V=dV/d\chi=0$ and $d^2 V/d\chi^2=m_\chi^2$ at
$\chi=f_\chi$.

The vacuum state of the Lagrangian at zero baryon number density
is defined by $U=1$ and $\chi= f_\chi$.  The fluctuations of the
pion and the scalar fields about this vacuum, defined through
\begin{equation}
U=\exp(i\vec{\tau}\cdot\vec{\phi}/f_\pi), \mbox{\ \ and \ \ } \chi
= f_\chi + \tilde{\chi}
\end{equation}
give physical meaning to the model parameters: $f_\pi$ as the pion
decay constant, $m_\pi$ as the pion mass, $f_\chi$ as the scalar
decay constant, and $m_\chi$ as the scalar mass.

The coupled classical equations of motion for the $U$ and $\chi$
fields of (\ref{lag-chi}) give rise to solitons for given winding
numbers corresponding to given baryon numbers with their structure
constrained by the classical scalar field. Nuclear matter is then
described classically by an FCC crystal and possibly a Fermi
liquid when quantized. When the system is strongly squeezed, there
is a phase transition from the FCC state to a half-skyrmion state
representing chiral restoration in dense system~\cite{LPRVchi}. It
is assumed that this transition remains intact when the initial
state is in Fermi liquid. Fluctuations on top of the background
crystal structure describe excitations of the pionic and scalar
quantum numbers, the properties of which are then governed by the
mean field values -- denoted $\chi^*$ -- of $\chi$ at a given
density fixed by the unit cell size $L$ of the crystal. What is
found in \cite{LPRVchi} is that the mean field of the scalar
$\chi$ mostly -- though not entirely -- governs the scaling
behavior of the parameters -- such as the pion decay constant --
of the Lagrangian indicative of the sliding vacuum structure. The
results are found to be in qualitative agreement with what is
found in BR scaling~\cite{BR91} as elaborated more precisely below
in NJL. The VM structure is not seen explicitly in the skyrmion
picture of \cite{LPRVchi} since there are no vector mesons in the
Lagrangian but it does not appear difficult to obtain it once
vector mesons are suitably implemented into the model Lagrangian.

\subsubsection{Nuclear matter from chiral symmetry decimation}
\itt Given an HLS/VM theory decimated to $\Lambda_{FS}$, how does
one go about obtaining nuclei and nuclear matter? The first thing
we need is a Fermi surface characterized by the Fermi momentum
$k_F\sim \Lambda_{FS}$ and the Fermi surface arises in effective
field theories as a quantum critical phenomenon. As sketched in
\cite{CND} and developed in \cite{LPMRV1,LPMRV2}, in the skyrmion
picture, a nucleus of mass number $A$ arises as a topological
soliton of winding number $W=A$. Nuclear matter is then given by
$W=\infty$. When quantized, the extended soliton system will
naturally possess a Fermi surface characterizing the filled Fermi
system.

In confronting nuclei and nuclear matter in nature, however, it is
more advantageous to work with explicit nucleon fields rather than
with multi-winding number skyrmions. When nucleons are explicitly
present in the theory, a nucleus will no longer emerge as a
topological object. Instead it must arise as a non-topological
soliton in a way conjectured by Lynn~\cite{lynn}. Nuclear matter
will then be more like a chiral liquid in Lynn's language. We
would like this soliton to emerge in a simple way from an
effective Lagrangian endowed with the parameters of the Lagrangian
{\it intrinsically density-dependent}. A systematic derivation of
such a soliton structure that is realistic enough is lacking at
the moment. However there is a short-cut approach to this and it
relies on Walecka's mean field theory of nuclear
matter~\cite{walecka}. The principal point we put forward is that
the Walecka mean-field solution of certain effective Lagrangian
(specified below) can be identified as the soliton -- topological
or non-topological -- solution described above.

Now, with the nucleon fields explicitly incorporated, there are
two (equivalent) ways to write down such an effective Lagrangian
of Walecka type that results from decimating down to
$\Lambda_{FS}$~\cite{BR:PR01}. One is the type-I approach in which
the heavy-meson degrees of freedom of the HLS Lagrangian are
integrated out and the other is the type-II one in which relevant
heavy-meson degrees of freedom are retained. The two versions give
equivalent descriptions of the same physics for the ground state
and low-frequency fluctuations.

The type-I Lagrangian has the form
 \be
\calL_I=\bar{N}[i\gamma_{\mu}(\del^{\mu}+iv^{\mu}
+g_A^\star\gamma_5 a^{\mu}) -M^\star]N -\sum_i C_i^\star
(\bar{N}\Gamma_i N)^2 +\cdots \label{leff}
 \ee
where the ellipsis stands for higher dimension and/or higher
derivative operators and the $\Gamma_i$'s Dirac and flavor
matrices as well as derivatives consistent with chiral symmetry.
The star affixed on the masses and coupling constants represents
the intrinsic $parametric$ dependence on density relevant at the
scale $\Lambda_{FS}$.  The induced vector and axial vector
``fields" are given by $v_\mu =
-\frac{i}{2}(\xi^{\dagger}\del_\mu\xi +\xi\del_\mu\xi^\dagger )$
and $a_\mu = -\frac{i}{2}(\xi^\dagger\del_\mu\xi
-\xi\del_\mu\xi^\dagger )$. In (\ref{leff}) only the pion ($\pi$)
and nucleon ($N$) fields appear explicitly: all other fields have
been integrated out. The effect of massive degrees of freedom that
are integrated out and that of the decimated ``shells" will be
lodged in higher-dimension and/or higher-derivative interactions.
The external electro-weak fields if needed are straightforwardly
incorporated by suitable gauging.

To write the Lagrangian for the type-II approach, we need to pick
the appropriate heavy degrees of freedom we want to consider
explicitly. This Lagrangian will be essentially the HLS Lagrangian
implemented with nucleon fields and chiral scalar heavy mesons
that are not taken explicitly into account. The relevant heavy
mesons for nuclear physics are a vector meson in the $\omega$
channel which is a chiral scalar and a flavor scalar $\sigma$ that
plays an important role in Walecka-type model for nuclear matter.
Assuming that $U(N_f)$ ($N_f=2$ for low-energy nuclear physics)
symmetry holds in medium, we can put the $\rho$ and $\omega$ in
the $U(N_f)$ multiplet and write a Harada-Yamawaki HLS Lagrangian
with the parametric dependence suitably taken into account. To
make the discussion simple, let us consider symmetric nuclear
matter in which case the type-II Lagrangian can be written in the
form of Walecka linear theory~\cite{walecka} with the parametric
density dependence represented by the star,
 \be \calL_{II} &=&
\bar{N}(i\gamma_{\mu}(\del^\mu+ig_v^\star\omega^\mu
)-M^\star+h^\star\sigma )N \nonumber\\ & &-\frac 14 F_{\mu\nu}^2
+\frac 12 (\partial_\mu \sigma)^2
+\frac{{m^\star_\omega}^2}{2}\omega^2
-\frac{{m^\star_\sigma}^2}{2}\sigma^2+\cdots\label{leff2} \ee
 where the ellipsis denotes higher-dimension operators. We should
stress that (\ref{leff2}) is consistent with chiral symmetry since
here both the $\omega$ and $\sigma$ fields are {\it chiral
singlets}. In fact the $\sigma$ here has nothing to do with the
chiral fourth-component scalar field of the linear sigma model
except near the chiral phase transition density; it is a
``dilaton" connected with the trace anomaly of QCD. The
possibility that this dilaton turns in-medium into the fourth
component of the chiral four-vector in the ``mended symmetry" way
\`a la Weinberg~\cite{weinberg-salam} near chiral restoration has
been discussed in \cite{beane}. Since we are far from the critical
density at which chiral restoration takes place, the vector mesons
are massive and the would-be Goldstone scalars (the longitudinal
components of the vector mesons) are absent. In (\ref{leff2}), the
pion and $\rho$-meson fields are dropped since they do not enter
in the mean field approximation but they can be put back if needed
(as for fluctuations in the pionic channel) in HLS symmetric way.
\subsection{Fermi-liquid decimation}\label{fermiliquid}
\itt Given the type-I (\ref{leff}) or type-II (\ref{leff2})
Lagrangians, the next step is to do the Fermi liquid decimation
(FLD). In \cite{schwenketal}, Schwenk, Brown and Friman build on
the effective interactions $V_{low-k}$ a decimation scheme to
arrive at Fermi-liquid parameters, namely the Landau effective
mass for the quasi-nucleon $m_N^\star$ and the quasiparticle
interactions ${\cal F}$. In doing this, the intrinsic
density/temperature dependence coming from the chiral symmetry
decimation was not included. It also makes the implicit assumption
that $n$-body interactions for $n>2$ are negligible. In any event,
it has been shown that both $m_N^\star$ and ${\cal F}$ are
fixed-point quantities. This approach purports to ``calculate" the
parameters of the ``penultimate" effective Lagrangian from first
principles.

An alternative method which we follow in this article is based on
the mapping of the effective field theory Lagrangians (\ref{leff})
and (\ref{leff2}) to the Fermi liquid fixed point theory as
beautifully explained in a review by Shankar~\cite{shankar}. The
mapping relies on the work done by Matsui~\cite{matsui} who showed
that {\it Walecka linear mean field theory is equivalent to Landau
Fermi-liquid theory}. Using this argument, it is then immediate to
map the mean-field solution of (\ref{leff}) or (\ref{leff2}) to
Landau EFT: In the mean field approximation, the two theories,
(\ref{leff}) and (\ref{leff2}), yield the same
results~\cite{gelmini,BR96}. This chain of reasoning, developed in
\cite{songetal,song,MR:MIGDAL}, has shown that the fluctuations on
top of the Fermi liquid fixed point -- represented by the mean
field of (\ref{leff}) or (\ref{leff2}) -- in response to EW
external fields can be related to the BR scaling parameter given
by the ratio $\Phi=m^\star/m$.
\subsection{The vector mass and gauge coupling
scaling}\label{d1-scaling}
\begin{figure}[hbt]
\vskip -0.cm
\centerline{\epsfig{file=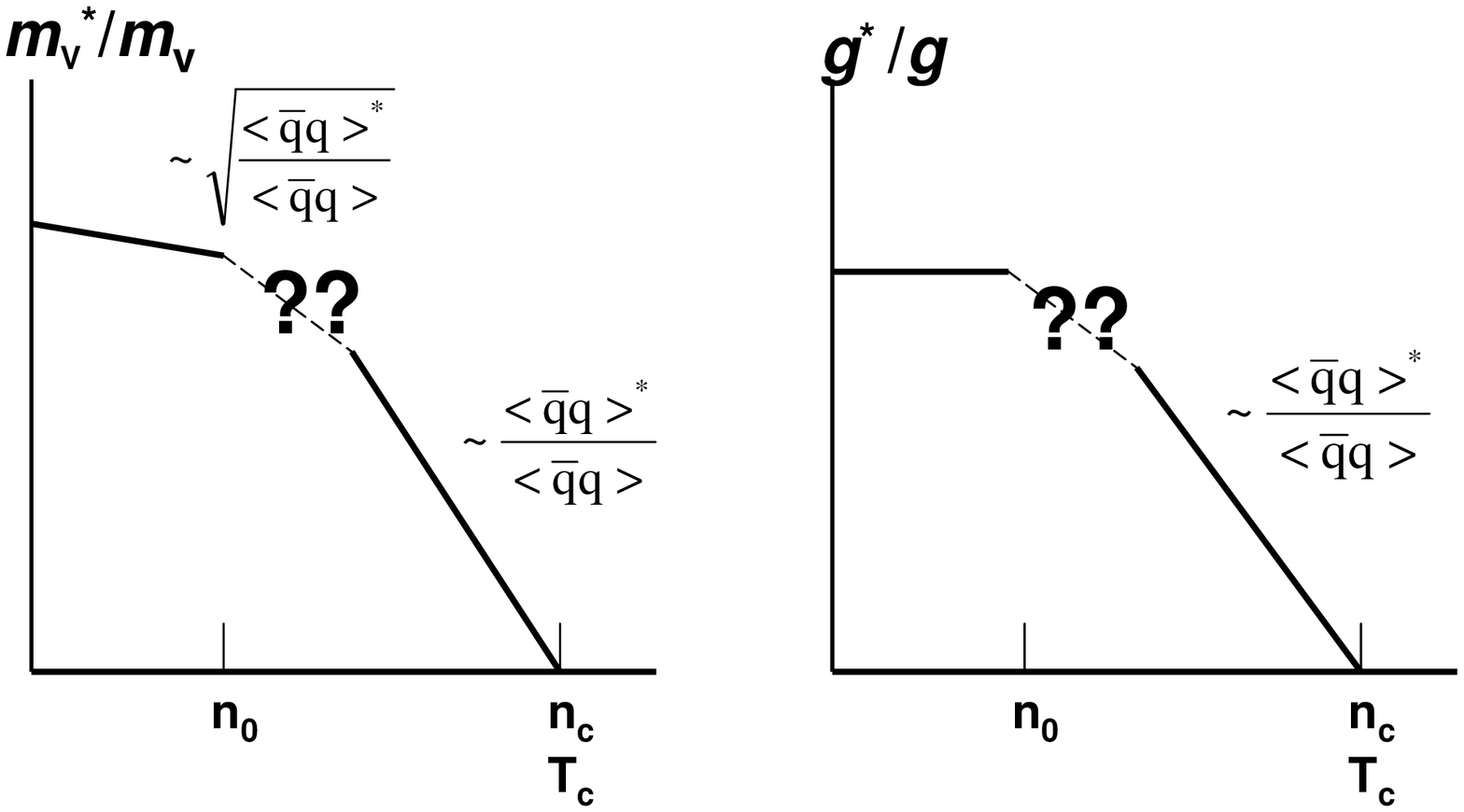,width=12cm,angle=-90}}
 \vskip -4cm
\caption{Schematic scaling behavior for $m_\rho^\star$ and
$g^\star$ vs. density. The cross-over region which is not
understood is marked by ??.}\label{double}
\end{figure}

 \itt In \cite{BR:PR01}, we discussed on
how the vector meson mass $m_V^\star$ and the gauge coupling
$g^\star$ in HLS theory with the vector manifestation -- both of
which are the essential elements of our EFT -- drop in medium as
one goes from below to above nuclear matter density. At very low
density, chiral perturbation theory provides relevant information
and near the critical density, the VM tells us how they scale. In
between, little could be said at the moment. Here we extrapolate
the scaling behavior from zero density (or temperature modulo
Dey-Eletsky-Ioffe low-temperature theorem~\cite{DEI}) to the
critical density (or temperature). Our picture is summarized in
Fig.\ref{double}.

Up to near nuclear matter density relevant to the chiral symmetry
decimation, the BR scaling of the mass is expected to go like
$\sqrt{\la\bar{q}q\ra^\star}$ and the vector coupling (more
precisely the HLS gauge coupling) to stay more or less unchanged.
The former is deduced from the GMOR relation for in-medium pion
mass which should hold in HLS theory for small density, with the
assumption that the pion mass does not change in density as
indicated, e.g., by recent $SU(2)_c$ lattice
results~\cite{muroya}. The latter generically follows from chiral
models and is consistent with the weakening tensor forces in
nuclei. We do not know precisely how the scaling changes as we go
above the matter density. However as we approach the critical
density, we learn from HLS theory matched to QCD that the vector
manifestation takes place, which means that both the hadronic mass
and the gauge coupling constant should scale linearly in
${\la\bar{q}q\ra^\star}$~\cite{HKR}. Thus near the chiral
transition point, $g^\star/m_V^\star$ should go to a constant of
density as required for quark number
susceptibilities~\cite{BR:PR96}.

\section{MEEFT or $More$ $Effective$ Effective Field
Theory}\label{eft}
\subsection{Doing effective field theory in nuclear physics}
 \itt
In this section, we consider the situation where one does not need
to go through the double decimation procedure described above.
Suppose one is interested only in describing low-energy
nucleon-nucleon scattering. For this, there are a variety of
equivalent ways to proceed~\cite{beaneetal}. First one writes down
an effective Lagrangian using the fields describing the degrees of
freedom relevant for the process in question. For instance, if one
is looking at the S-wave two-nucleon scattering at low energy, say
much less than the pion mass, one can simply take only the nucleon
field as an explicit degree of freedom, integrating out all
others, including the pion. The pions which are important due to
the broken chiral symmetry may be introduced perturbatively. A
systematic power counting can then be developed and used, in
conjunction with an appropriate regularization (e.g., the power
divergence substraction (PDS) scheme~\cite{beaneetal}), to compute
the scattering amplitude by summing to all orders a particular set
of diagrams. In principle, the parameters of the pionless
Lagrangian could be determined for a given scale by lattice QCD.
In this sense, such a systematic higher-order calculation in this
EFT can be considered to be $equivalent$ to doing QCD. In
practice, however, the parameters are obtained from experiments.
For two nucleon systems, two classes of parameters are to be
determined from experiments. One is single-particle vertex and
this is given by on-shell information. The other is intrinsically
two-particle in nature requiring data on two-nucleon processes.
When the parameters are fully determined, this procedure with the
pionic contributions taken into account perturbatively does lead
to sensible results. For instance, it correctly reproduces such
standard nuclear physics results as the effective-range formula.

The EFT which adheres strictly to order-by-order consistency in
the power counting (that we shall refer to as ``purist's EFT")
however suffers from the lack of predictivity. The number of
unknown parameters increases rapidly as the number of nucleons
involved in the process increases. Thus even if two-nucleon
systems are well described by the EFT in question, treating
systems involving more than two or three nucleons becomes
prohibitively difficult, if not impossible. Furthermore treating
pions as perturbative misses the power of the ``chiral filter
mechanism"~\cite{KDR} that plays an important role in predictive
calculations~\cite{PMR}. At present, going to nuclear matter is
out of reach by this EFT approach.

Even if such a method were available so that we could write down
an effective Lagrangian on top of a Fermi sea, the parameters of
the penultimate effective theory for, say, nuclear matter, would
be very far from the first principles, QCD: Calculating them would
be somewhat like calculating the boiling point of the water
starting from QED. We propose here how to circumvent this impasse.
Our approach proposed here is admittedly indirect and drastically
simplified. Broadly speaking, it involves a two-step decimation
starting from the chiral Lagrangian that is matched to QCD at the
chiral scale $\Lambda_\chi\sim 1$ GeV. The key ingredient that
allows this feat is the highly refined ``standard nuclear physics
approach (SNPA)" and the objective is to marry the SNPA to an EFT.

As an illustration of our strategy, we briefly discuss here how
the marriage can be effectuated. As summarized
recently~\cite{MEEFT,BR:PR01}, a thesis developed since some time
posits that by combining the standard nuclear physics approach
(SNPA) based on potentials fit to experiments with modern
effective field theory, one can achieve a more predictive power
than the {\it purist's EFT} alone can. The idea was recently given
a test in a variety of electroweak processes in nuclei, in
particular, the solar $pp$ fusion and $hep$ processes where highly
accurate predictions could be made free of
parameters~\cite{park-pp-hep}. As alluded above, a closely related
mechanism has been found at work in nuclear effective interactions
that figure in nuclear structure calculations~\cite{bogneretal}.
We briefly sketch these two recent developments which are closely
connected to the issue at hand.
\begin{figure}[tb]
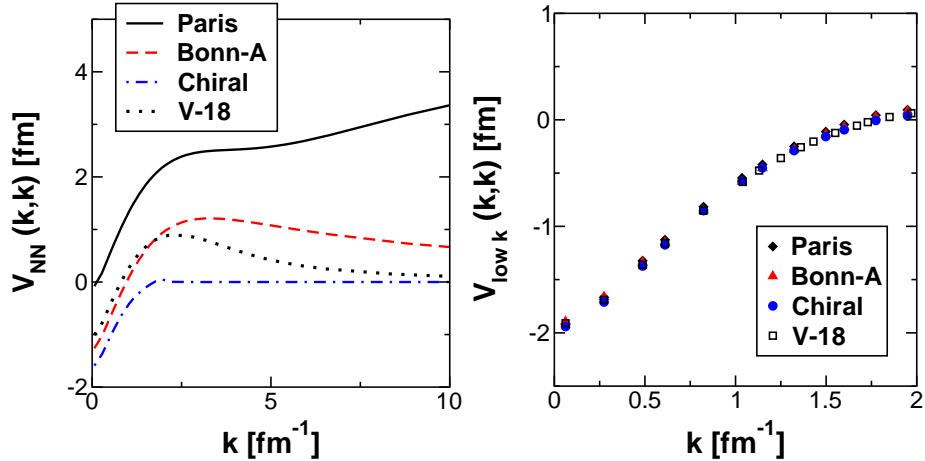

\vskip 0cm
 \centerline{\epsfig{file=vk1.eps,width=6cm,angle=0}
 \epsfig{file=vk2.eps,width=6cm,angle=0}}
 \vskip 0cm
\caption{Diagonal matrix elements of several ``realistic
potentials" $V_{bare}$ and the effective potentials $V_{low-k}$ as
a function of relative momentum in the $^1$S$_0$ partial wave. The
effective potentials are calculated with the cutoff $\Lambda=2$
fm$^{-1}$. From ~\cite{bogneretal}.} \label{bogneretal}
\end{figure}

\subsection{Effective interactions in nuclei}
\itt The calculation of the EW response functions that will be the
subject of the next subsection relied on an implicit assumption on
the effective forces that enter in the calculation of the wave
functions. A recent development provides a support to this
assumption.

Consider two nucleon scattering at very low energy, a process very
well understood in SNPA. The relevant $T$ matrix for the
scattering is a solution of the Lippmann-Schwinger equation with a
``bare" potential $V_{bare}$ figuring as the driving term. The
long-range part of the bare potential $V_{bare}$ is governed by
chiral symmetry, namely, by a pion exchange and hence is
unambiguous. But the short-range parts are not unique. Even if the
potentials are determined $accurately$ by fitting scattering data
up to say $E_{lab}\sim 350$ MeV, those potentials that give the
equivalent phase shifts can differ appreciably, in particular in
the short-distance parts. In terms of effective field theory, what
this means is that while the long-range parts given by
``low-order" expansion are the same for all the realistic
potentials, shorter-range terms that are given by higher order
terms can differ depending upon how they are computed. They will
depend upon how the power counting is organized, what
regularization is used etc. In practice, those ``higher-order"
terms are fixed by fitting to experimental data. The examples for
such realistic potentials are the Paris potential, the Bonn
potential, the chiral potential etc. Suppose in the integral
equation satisfied by the scattering $T$-matrix, one integrates
out the momentum scales above a given cutoff $\Lambda$
sufficiently high to accommodate the relevant degrees of freedom
and probe momentum but low enough to exclude the massive scales
that do not explicitly figure in the theory, with the requirement
that the resulting effective potential reproduce the phase shifts
while preserving the long-range wave function tails as given by
the half-on-shell (HOS), $T(k^\prime,k;k^2)$. To the extent that
the HOS $T$ matrix is a physical quantity, it should be
independent of where the cutoff is set
 \be
\frac{d}{d\Lambda} T(k^\prime,k;k^2)_\Lambda=0.
 \ee
This condition leads to a renormalization group equation (RGE) for
the effective potential, denoted $V_{low-k}$. It is important to
note that the fit to experiments {\it defines the complete theory}
for the probe momentum $k\ll \Lambda$.  Now since the HOS $T$ is
fit to experiments, integrating out the momentum component
$p>\Lambda$ transfers the physics operative above the cutoff into
the counter terms that are to be added to the bare potential to
give the effective one, $V_{low-k}$. Bogner et
al~\cite{bogneretal} have shown that the resulting $V_{low-k}$ is
independent of the bare potential one starts with as long as it is
consistent with the chiral structure, i.e., the long-range tails
of the wave functions, and the $T$ is fit to experiments. It is
important to recognize that this strategy is none other than that
of the SNPA.
\begin{figure}[htb]
\vspace{-8mm}
\begin{center}
\includegraphics[scale=1.15,clip=]{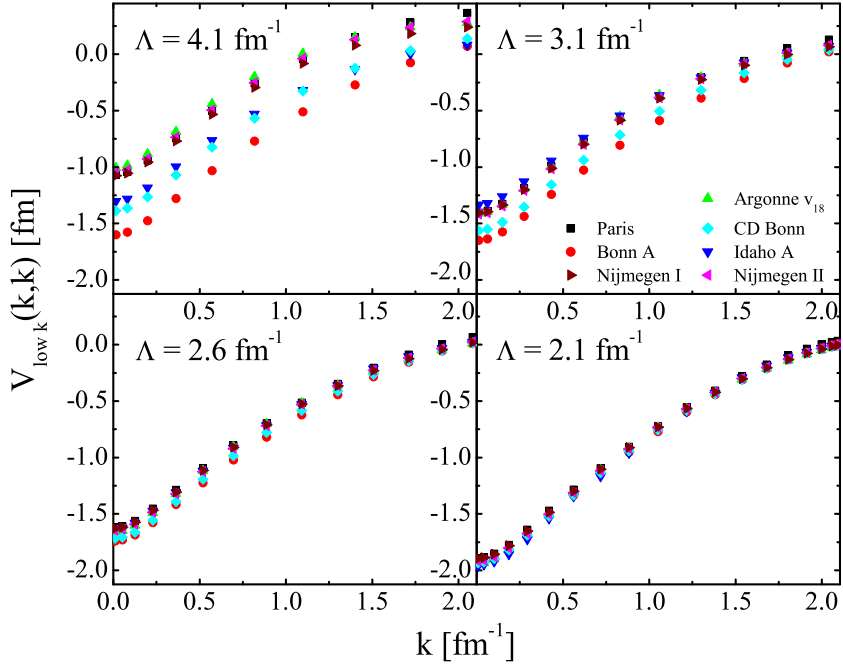}
\end{center}
\vspace{-8mm} \caption{The collapse of the diagonal momentum-space
matrix elements of $V_{low-k}$ as the cutoff is lowered to
$\Lambda = 2.1 \, {\rm{fm}}^{-1}$ in the $^3$S$_1$ partial wave.}
\label{collapse}
\end{figure}

The results of Ref.\cite{bogneretal} are reproduced in
Fig.\ref{bogneretal}. The figure on LHS shows the variety of bare
potentials that are fit to experiments and the one on the RHS the
effective potentials collapsing into a universal curve at
$\Lambda=2$ fm$^{-1}$ after the integrating-out procedure. How the
collapse occurs from different $V_{low-k}$'s for different
potentials is shown in Fig.\ref{collapse}. For $\Lambda >$ 2
fm$^{-1}$, $V_{low-k}$'s can be different for different potentials
but they ``collapse" to one universal curve for $\Lambda=2.1$
fm$^{-1}$. The reason for this is easy to understand: that the
phase shift analyses from which the two-body interactions in Fig.
\ref{collapse} were fit were carried out for experiments up to
laboratory energies of $\sim$ 350 MeV, which corresponds to a c.m.
momentum of 2.1 fm$^{-1}$~\footnote{At about this energy
substantial inelasticity sets in, complicating analysis.}.
Although the phase shifts were determined on-shell, as we noted in
Section \ref{intro} the interaction is not far off-shell when used
in nuclei. This is how one can understand that the half-off-shell
$V_{low-k}$ reproduced the same diagonal matrix elements for the
various potentials fit to experiments to the given cut-off scale.
The point is that off-shell effects are unimportant in
considerations of the smooth parts of the shell-model wave
functions. Of course, these effects can be large if the wave
functions are forced far off shell by short-range two-body
interactions, but these will involve a scale higher than our 2.1
fm$^{-1}$ not constrained by the experimental data.

That while the bare potentials are widely different for all
momenta, all realistic potentials give the identical effective
potential illustrates the power of the strategy that {\it combines
the accuracy achieved by the SNPA and chiral effective field
theory of QCD into a ``more effective effective field theory
(MEEFT)."}  The key lesson from this result is that the
short-distance part of the potential which represents higher
orders in EFT power counting schemes which differ for different
counting schemes may be different from one ``realistic potential"
to another but when suitably {\it regularized taking into account
the constraints by experimental data}, the resulting effective
potential comes out unique. For instance, the ``chiral potential"
which is consistent with the chiral counting, and hence presumed
to be more in line with the tenet of EFT and the $v_{18}$
potential which, apart from the long-range part, is not, can
differ at ``higher orders" but gives the same $V_{low-k}$. This
reflects how the MEEFT works.

An identical mechanism is at work for the weak matrix elements
relevant for the solar neutrino processes~\cite{park-pp-hep} as
described below.

The $V_{low-k}$ is the basis of nuclear structure calculations
replacing the role of G-matrix. For low-energy processes, it is
insensitive to short-distance physics properly taking into account
the standard short-range correlations. It is also the input for
field theoretic calculation of the Landau parameters for nuclear
matter~\cite{schwenketal} which are the Fermi liquid fixed point
quantities.
\subsection{Electroweak processes in nuclei}
\itt The next question we raise is: How can one do a unified
effective field theory calculation which is truly predictive for
the following processes?
 \be
p+p&\rightarrow& d +e^+ +\nu_e,\label{pp}\\
^3{\rm H}&\rightarrow& ^3{\rm He} +e^- +\bar{\nu}_e,\label{beta}\\
p + ^3{\rm He}&\rightarrow& ^4{\rm He} + e^+ +\nu_e.\label{hep}
 \ee
The processes (\ref{pp}) and (\ref{hep}) take place in the Sun,
playing an essential role for the solar neutrino problem but have
not been measured in the laboratories while the process
(\ref{beta}) is accurately measured.  Given the triton beta decay
data, the objective then is to make an accurate prediction for the
rates for (\ref{pp}) and (\ref{hep}). This highly non-trivial feat
has been accomplished recently by Park et al~\cite{park-pp-hep} by
an MEEFT. We note that such a predictive calculation is not yet
feasible in other EFT strategies developed so far (e.g., the
purist's EFT).

The MEEFT strategy of \cite{MEEFT} employed in the publications by
Park et al.~\cite{park-pp-hep} goes as follows. First one picks a
potential fit to on-shell NN scattering up to typically
$E_{lab}\sim 300$ MeV plus many-body (typically three-body)
potentials, the combination of which is to describe accurately
many-nucleon scattering data. As stated, this potential is
required to be consistent with chiral symmetry, the symmetry of
QCD, which means that the long-range part of the potential is
dictated by the pion exchange. An example for such a potential is
the $v_{18}$ potential which has been used in the numerical
calculation. Next one computes the weak current in a chiral chiral
perturbation theory, e.g., heavy-baryon chiral perturbation
theory, to an appropriate order in the power counting, say, $Q^n$
where $Q$ is the characteristic momentum scale. The current matrix
element is then computed with this current and the ``realistic"
wave functions computed with the given potential. In doing this,
one integrates out all momentum components above the cutoff
$\Lambda$, the effect of which is a set of counter terms appearing
in the effective current. Working to the next-to-next-to-next
order (i.e., $Q^3$) relative to the leading matrix element given
by the single-particle Gamow-Teller matrix element in
(\ref{pp})-(\ref{hep}), one encounters a number of counter terms
but symmetry consideration reduces them to a single combination,
denoted $\hat{d}^R$ in \cite{park-pp-hep}, entering in all three
processes. The counter term coefficient $\hat{d}^R$ is
cutoff-dependent and so is the matrix element of the current with
the momentum component $p>\Lambda$ integrated out. The numerical
values of these individual terms differ from one potential to
another but the sum of the two does not. The strategy then is to
exploit that the short-range component of the interactions must be
$universal$, that is, independent of the mass number. This means
that the coefficient $\hat{d}^R$ can be determined for any given
cutoff from the accurately measured process (\ref{beta}). Since
there are no other unknowns to order $Q^3$, we have a parameter
free theory to calculate all the matrix elements that figure in
(\ref{pp}) and (\ref{hep}). Indeed this is what was done in
\cite{park-pp-hep}: The $S$ factor for the $pp$ fusion process was
calculated within the accuracy of 0.4\% and that for the $hep$
process within the accuracy of 15\%. Both results, particularly
the latter, are of unprecedent accuracy unmatched by other
calculations~\footnote{We remind the readers that up to date, the
calculated values for the $hep$ $S$ factor varied by orders of
magnitude, hence completely unknown.}.

\section{Saturation and Dirac Phenomenology}\label{diracpheno}
\itt In detail the situation with dropping masses is complicated
because the pion, both in lowest order and in two-pion exchange,
contributes to the energy, but $m_\pi$ is most likely unscaled,
whereas the scalar and vector mesons, two most important effective
fields in nuclear medium, are. Rapp et al~\cite{RMDG} showed that
given these scalings, saturation does come at the right density.
The chief contributions to the binding energy come from the
$\sigma$ and $\omega$ exchange. Here we show schematically how
these behave in the ``swelled" world.

Consider a Hamiltonian
 \be
H=\sum_i\frac{1}{2m_N}\nabla_i^2 +\frac 12\sum_{ij} (g_\omega
\frac{e^{-m_\omega r_{ij}}}{r_{ij}}-g_\sigma\frac{e^{-m_\sigma
r_{ij}}}{r_{ij}})
 \ee
with ground state energy
 \be
H\psi=E_0\psi.
 \ee
Now scale all $m$'s
 \be
m_N^\star=\lambda m_N, \ m_{\sigma,\omega}^\star=\lambda
m_{\sigma,\omega}.
 \ee
Then
 \be
H^\star (\lambda, \vec{x})=\lambda H(1, \vec{r})
 \ee
where
 \be
\vec{x}=\lambda \vec{r}.
 \ee
Now
 \be
H^\star (\lambda, r)\psi=E_0(\lambda)\psi
 \ee
or
 \be
\lambda H(1, x)\psi=E_0 (\lambda)\psi.
 \ee
But $H(1,x)$ is just a relabelling of $H$ via $r\rightarrow x$,
etc. so the solution of
 \be
H(1,x)\psi=E\psi
 \ee
for the lowest eigenvalue is $E_0$, or
 \be
E_0=E_0(\lambda)/\lambda.
 \ee
Thus
 \be
E_0 (\lambda)=\lambda E_0,
 \ee
and the scaled system (with $\lambda$) is less bound than the
original unscaled system since $\lambda=0.8$ at nuclear matter
density. The replacement of the correlated two $\pi$ exchange,
with the two $\pi$'s in a relative S-state by a $\sigma$ with
scaling mass is, however, not a good approximation as shown by
Rapp et al~\cite{RMDG}. Even with the loss in binding energy,
because of the scaling, saturation does not occur in the correct
region of densities, as can be seen from Fig.2 of the quoted
paper. Although crossed channel exchange of a $\rho$ with
decreased mass increases the attraction, such a decreased $\rho$
mass increases the (repulsive) part of the Lorentz-Lorenz
correction of the pion coupling to $NN^{-1}$ and $\Delta N^{-1}$
bubbles. In addition, the form factor $\Lambda_\rho^*$ of the
t-channel $\rho$ exchange must be scaled. Also repulsive contact
interactions in the $\pi\pi$ scattering needed to preserve chiral
invariance are proportional to ${f_\pi^*}^{-2}$; they balance the
increase in attraction from t-channel $\rho$-exchange. When all of
the constraints from chiral invariance and BR scaling are
enforced, the 2$\pi$ exchange potential -- which ar low densities
behaves approximately like a scalar meson with BR-scaling mass --
gives an effective scalar interaction corresponding to a $\sigma$
in which the decrease in mass is slowed down. This decrease in the
rate of dropping enables saturation at the correct density. From
the references in \cite{RMDG} the interested reader will be able
to construct the unphysical artifacts that arise from neglecting
the chiral constraints in the two-pion exchange interactions and
convince oneself that they are necessary on physical as well as on
formal grounds.

Note that with BR scaling, and the above modifications of it, the
mechanism of saturation is quite different from that of
introduction of three-body forces, as in Pieper et
al~\cite{vijay}, who carry out an essentially exact calculation of
binding energy of light nuclei in the conventional approach with
bare hadron masses (that does not $explicitly$ implement the {\it
intrinsic density dependence} required by matching to QCD). We
suggest that these may not be relevant to the physical situation.

From the foregoing, we see that the dynamics in the ``sliding
vacuum" substantially changed the binding energy, but in many
other respects is only slightly different from that in the
perturbative vacuum at low densities. Indeed, now it is clear why
Brown and Rho found in their early work the change in tensor force
as indicative of dropping masses; there the $\rho$-meson, which
does drop in mass, beats against the $\pi$-meson, which does not.

\begin{figure}
\vskip -2cm
\centerline{\epsfig{file=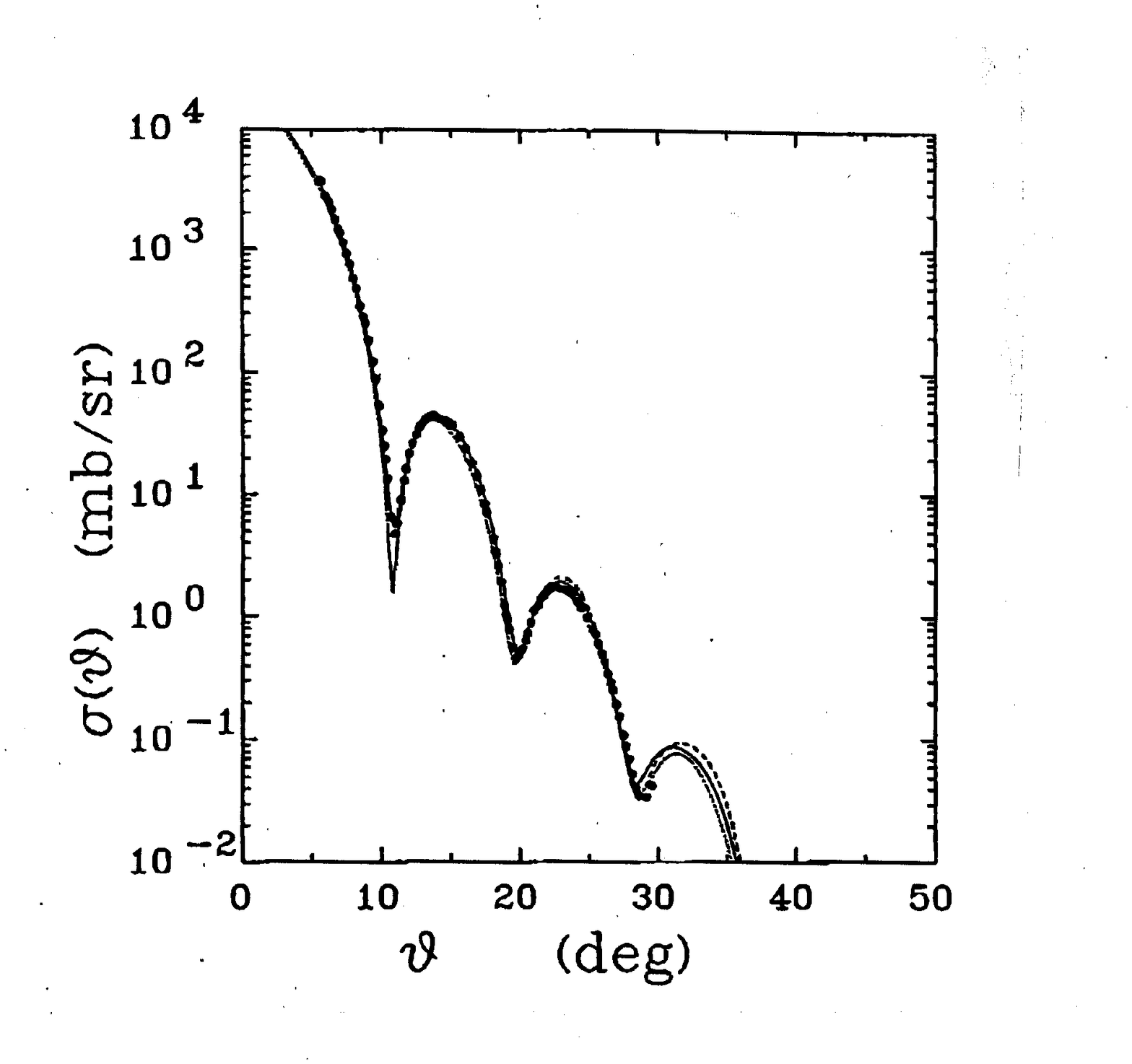,width=14cm,angle=-90}}
 \vskip -1cm
 \caption{Results of Tjon and Wallace (see Brown, Sethi and
 Hinz~\cite{hintz}) for the differential cross
section of 500 MeV protons scattered elastically from $^{40}$Ca.
The dotted line gives the IA2 results. For the solid curve a
linear scaling with density was assumed, with $m_\sigma^\star
(n_0)/m_\sigma=m_\omega^\star (n_0)/m_\omega=m_\rho^\star
(n_0)/m_\rho=0.85$. Note that the two are essentially on top of
each other. For the dashed line $m_\sigma^\star$ and
$m_\omega^\star$ scaled in this way but $m_\rho$ was held constant
at its free-space value. The experimental points are given as
solid round dots.}\label{fig1}
\end{figure}

\begin{figure}[ht]
\vskip -2.cm
\centerline{\epsfig{file=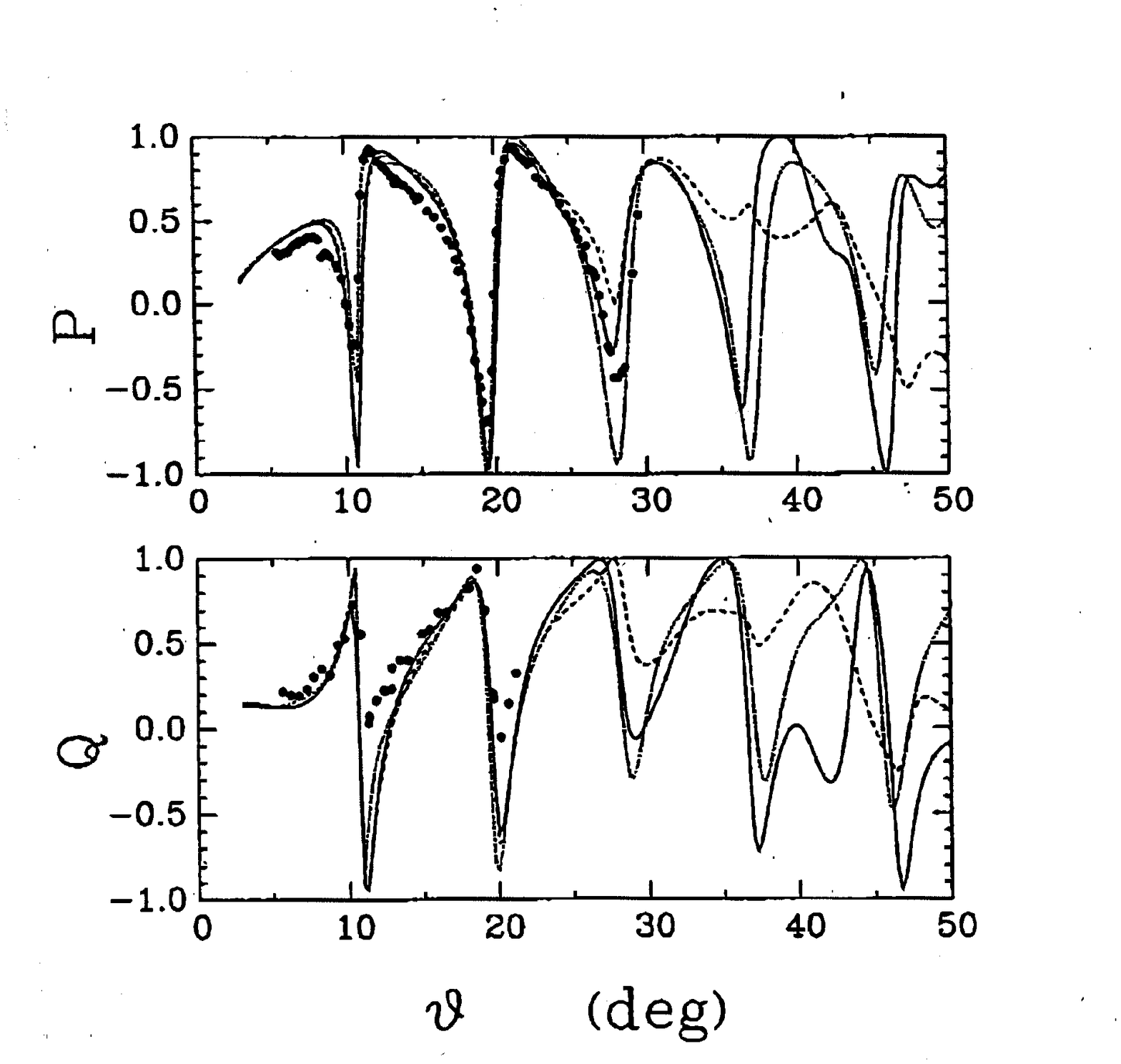,width=14cm,angle=-90}}
 \caption{Results of Tjon and Wallace for the spin observables for
500 MeV protons scattered elastically from $^{40}$Ca. The lines
represent the same as in Fig.\ref{fig1}. Note that the dashed line
starts deviating strongly from the others beyond about 30 degrees.
}\label{fig2}
\end{figure}

It is quite remarkable that in much of nuclear structure, it does
not seem to matter whether BR scaling is operative or not. For
instance, consider one of the most successful predictions in
nuclear physics, the Dirac phenomenology. The most complete
theoretical paper on this was by Clark, Hama and
Mercer~\cite{clarketal}. Predictions of this theory as compared
with experiment were incredibly successful, the rapid oscillation
in the polarization and spin rotation as function of angle being
reproduced in great detail. Brown, Sethi and Hintz~\cite{hintz}
persuaded John Tjon and Steve Wallace to put a linear scaling of
meson masses with density into their relativistic impulse
approximation (RIA) calculation. We reproduce Figs. 12 and 13 from
Brown, Sethi and Hinz as Fig.\ref{fig1} and Fig.\ref{fig2} here.

Looking at these figures with a magnifying glass, one can see that
the fit to data is not worsened by putting in density-dependent
masses~\footnote{As noted~\cite{BR:qm02}, the vector coupling and
associated constants are presumed to be non-scaling up to nuclear
matter density.}. We now outline the calculations that Tjon and
Wallace have carried out, scaling meson masses other than the pion
as
 \be
m_M^\star/m_M\approx 1-0.15n/n_0
 \ee
and nucleon mass as
 \be
m_N^\star/m_N\approx 1-0.2 n/n_0.
 \ee
This scaling is somewhat milder than what we now advocate
$m_M^\star/m_M\approx 1-0.2n/n_0$ and $m_N^\star/m_N\approx
1-0.3n/n_0$. The Tjon-Wallace calculation, putting in the local
density, was incredibly complicated, with 128 components of Dirac
wave functions.  Tjon ran part of the program, Wallace the rest,
and they had to be together to complete it. Such a daunting
calculation is highly unlikely to be repeated in the foreseeable
future. So we will draw what conclusion we can draw from their
results.

Before proceeding, we must point out several caveats here in
re-interpreting the Tjon-Wallace results in terms of our language.

Whereas the IA2 used by Tjon and Wallace to check the scaling
masses in Dirac phenomenology, with results given in Brown, Sethi
and Hintz did not include the effective mass $m_N^\star$, it did
keep positive and negative energy states in a plane wave
decomposition, and handled pair theory correctly. As shown in
\cite{BWBS,machleidt}, in perturbation theory this is equivalent
to use of a nucleon effective mass in Walecka mean field theory.

The $m_N^\star$ obtained in the above way by Tjon and Wallace was
$\sim 0.8 m_N$ at $n=n_0$, somewhat less than the $0.7m_N$ that we
advocate. Also we now have the scaling meson masses as $\sim
0.8m_M$ at $n=n_0$. Thus their $m_N^\star$ dropped $\sim 4/3$
times faster than the scaling meson masses. The linear scaling
was, however, consistent with present theory, which as we return
to later, has $m_N^\star$ scaling $3/2$ times faster than
$m_M^\star$.

From the Tjon-Wallace results, we conclude that Dirac
phenomenology is preserved well in all detail with the
approximately linear scaling. In other words their fine structure
of the polarization variables survives not only BR scaling, but
even scaling with effective nucleon mass scaling somewhat faster
than the meson masses. There is no discernable effect of the
``breaking" of this scaling by the pion effective mass being kept
constant. We thus find unfounded those claims that since
conventional nuclear theory provides a successful description of
the nuclear many-body problem, changes such as BR scaling should
not be made in that it would upset the successful predictions. We
have shown that much of nuclear structure -- with a few notable
exceptions which can be understood -- are inert to the change of
mass parameters.
\section{In-Medium Pion Decay Constant}\label{axialcharge}
 \itt
The fact that most of nuclear structure physics do not ``see" the
BR scaling does not mean that $all$ nuclear observables are inert
to it. We have mentioned that the tensor force in nuclei has the
scaling $\rho$ meson beating against the unscaling pion, which
should leave a distinct imprint. In this section we discuss
several observables that probe the scaling $f_\pi^*/f_\pi$, the
bona-fide order parameter of chiral symmetry in the hadronic
language.

In section \ref{parametric}, the notion of parametric density
dependence in HLS theory was described. Now when such HLS
Lagrangian is applied to nuclear matter, the intrinsic density
dependence factor $\Phi$ can in turn be related to a Landau
parameter which addresses many-body interactions as described in
section \ref{fermiliquid}, an indication that the change of QCD
vacuum in medium is intricately tied to many-body properties. This
demonstrates that it is most likely to be futile to try to
separate what one would attribute to QCD from what amounts to
many-nucleon dynamics.

Consider long wavelength fluctuations on top of the ``vacuum"
defined by such parameters. As effective degrees of freedom, we
may pick the pions and the nucleons and integrate out vector
mesons and other heavy hadrons including scalars from the HLS
Lagrangian. The resulting Lagrangian will take the same chiral
symmetric form as in the free space except that the parameters of
the Lagrangian intricately depend on density. With this
Lagrangian, the power counting will be formally the same as in
ChPT based on the Lagrangian whose parameters are defined in
matter-free space. According to the discussion given in Section
\ref{d1-scaling}, we have, up to near nuclear matter density,
 \be
\Phi (n)\equiv m_V^*/m_V\approx F_\pi^* (n)/F_\pi\approx
\left(\la\bar{q}q (n)\ra^*
/\la\bar{q}q\ra\right)^{1/2}\label{condensate}
 \ee
where $F_\pi$ is the bare (parametric) pion decay constant. The
(last) relation between the pion decay constant and the quark
condensate comes from the Gell-Mann-Oakes-Renner mass formula for
the pion in medium. The in-medium quark condensate can be
estimated for low density in terms of the pion-nucleon sigma term
$\Sigma_{\pi N}$ using the low-energy
theorem~\cite{drukarev,cohen}
 \be
\la\bar{q}q (n)\ra^* /\la\bar{q}q\ra =1-\frac{\Sigma_{\pi
N}}{m_\pi^2 f_\pi^2}n+\cdots
 \ee
Using the presently accepted value for the sigma term $\Sigma_{\pi
N}\approx 45$ MeV, we get
 \be
\la\bar{q}q (n)\ra^* /\la\bar{q}q\ra\approx
(1-0.36n/n_0)^{1/2}.\label{br1}
 \ee
which gives $\la\bar{q}q (n_0)\ra^* /\la\bar{q}q\ra\approx 0.8$ at
nuclear matter density. On the other hand, the quantity $\Phi$ of
(\ref{condensate}) can be obtained by relating the vector-meson
mass scaling to the Landau parameter $F_1$ and then using the
relation between the Landau parameter and the anomalous
gyromagnetic ratio of heavy nuclei as well as the properties of
nuclear matter~\cite{songetal,MR:MIGDAL}. One gets
 \be
\Phi (n)\approx (1+0.28n/n_0)^{-1}\label{Phi}
 \ee
where $n_0$ is normal nuclear matter density. At nuclear matter
density, this gives $\Phi(n_0)\approx 0.78$. Thus the two
relations (\ref{br1}) and (\ref{Phi}) agree up to nuclear matter
density. In what follows we shall adopt (\ref{Phi}) since it
directly reflects BR scaling. Furthermore the model-independent
relation (\ref{br1}) depends sensitively on the value of the $\pi
N$ sigma term, the precise value of which is still highly
controversial and uncertain. For a recent summary, see
~\cite{sigmaterm}.

Given (\ref{Phi}), one can make a few independent predictions. The
first quantity we can look at is  Warburton's
$\epsilon_{MEC}$~\cite{warburton} for axial-charge transitions in
heavy nuclei. Using the chiral Lagrangian with the parametric
dependence defined above, one can easily compute this in the
leading chiral order~\footnote{According to the chiral filter, the
soft-pion term should dominate with next-order corrections
suppressed for the axial-charge transitions.}, that is, in tree
order~\cite{kuboderarho}. The prediction is that~\cite{BR:PR01}
 \be
\epsilon_{MEC} (n=n_0/2)&\approx& 1.63,\nonumber\\
 \epsilon_{MEC} (n=n_0)&\approx& 2.02.
 \ee
These should be compared with the observed values
$\epsilon_{MEC}^{exp}=1.60\pm 0.05 (2.01\pm 0.10)$ for the mass
numbers $A=50 (208)$.

The next thing we consider is the recent experiment on deeply
bound pionic atoms~\cite{geissel,yamazaki}. There is an on-going
discussion as to whether this experiment is signalling ``partial
restoration" of chiral symmetry. There can be a variety of ways to
approach this problem. In the framework we are adopting here with
the chiral Lagrangian based on HLS/VM, what the experiment
provides is an information on the only scale dependent parameter
in the theory, namely, the ratio $F_\pi^* (n)/F_\pi$ at $n\lsim
n_0$. In the tree order, this ratio is given by (\ref{Phi}) which
at nuclear matter density is
 \be
(F_\pi^* (n_0)/F_\pi)^2=0.61\label{prediction}
 \ee
with a theoretical uncertainty of $\sim 10\%$ inherent in nuclear
$\delta g_l$. This agrees with the value extracted -- in the tree
order -- from the pionic atom data~\cite{yamazaki}
 \be
(F_\pi^* (n_0)/F_\pi)^2=0.65\pm 0.05.
 \ee

One can understand this result obtained in the leading order ChPT
with HLS/VM in a way more familiar to nuclear physicists, as
follows based on the works by Friedman~\cite{friedman} and Weise
and coworkers~\cite{weiseatom,weise}.  For instance, Friedman
invokes density dependence in two places. One is to incorporate
the ratio~\cite{weise} $f_\pi^\star/f_\pi \approx
\sqrt{\la\bar{q}q\ra^\star/\la\bar{q}q}\approx 0.78$ at $n=n_0$
which enters into the Weinberg-Tomozawa isovector term in the
pion-nucleon scattering amplitude~\footnote{This actually goes
back to the work of Lutz, Klimpt and Weise~\cite{lutzetal} for the
value of $f_\pi^\star$ obtained from the Gell-Mann-Oakes-Renner
relation assumed to hold in matter with the pion mass unscaled by
density. At low density, we expect this assumption to be valid. At
present, the theoretical evidence for an unscaled pion mass comes
from two-color ($SU(2)_c$) QCD on lattice~\cite{muroya} although
the error bars are a bit too big to confirm the constancy and of
course the two-color QCD may not reflect the real QCD and from a
skyrmion description~\cite{LPRVchi}.

Friedman invokes the in-medium pion decay constant expressed in
terms of the pion-nucleon sigma term. As noted above, the
numerical value turns out to be equivalent to $\Phi (n_0)$ used in
this paper. The point we want to stress is that this quantity is
the one that figures in BR scaling.}. This scaling can be
identified~\cite{BR:qm02} as the ``intrinsic" density dependence
$required$ by matching to QCD. The other medium modification
Friedman needs goes back to the relativistic impulse approximation
used by Birbrair and collaborators~\cite{birbair} which gives
small components of the Dirac wave functions for the nucleon
enhanced by the factor
 \be
F=m_N/M (r)
 \ee
where $M (r)=m_N +\frac 12 [S(r)-V(r)]$ with $S(r)$ and $V(r)$ the
vector mean fields. In fact, if we write
 \be
M(r)=m_N+S-\frac 12 (S+V)
 \ee
and note that up to nuclear matter density $(S+V)/2$ turns out to
be only about 10\% of the $S$ in magnitude~\footnote{This
observation is the basis of the suggestion made in ~\cite{weise2}
that fluctuations in nuclear matter be computed around the
``shifted vacuum" at which $(S+V)/2$ is exactly equal to zero.
This ``vacuum" at this point is close to the Fermi-liquid fixed
point. Note that this is analogous to the notion of fluctuating
around BR's ``sliding
vacuum"~\cite{songetal,LPRVchi,song,MR:MIGDAL}.}, we see that
$M(r)\simeq m^\star (r)$, the nucleon effective (Landau) mass.

In Brown-Rho (BR) scaling~\cite{BR91}, the nucleon effective mass
scales more rapidly than $m_\rho^\star$ or $f_\pi^\star$ because
of pionic effects in the solitonic background going as
$\sqrt{g_A^\star/g_A}$. In the original formulation in terms of
skyrmions~\cite{BR91}, this factor entered when the Skyrme quartic
term was taken into account for $g_A$. Now the Skyrme quartic term
is known to contain a lot more than what naively appears to result
when heavy degrees of freedom, e.g., all heavy excitations of the
$\rho$ vector meson quantum numbers, are integrated out. It seems
to represent physics ranging from part of pionic effects to
$extreme$ short-distance effects (e.g., proton
decay~\cite{rubakov}). (For discussions on some of these matters,
see \cite{CND}.) On the other hand, one can also calculate its
effect from matching of chiral Lagrangian theory with sliding
vacuum to Landau Fermi liquid theory as discussed in
\cite{songetal,song,MR:MIGDAL}. There this factor arose as a
pionic contribution correlated with heavy modes in the
vector-meson channel, given by the formula $g_A^\star/g_A=(1-
\frac 13 \tilde{F}_1 (\pi)\frac{m_V^\star}{m_V})^{-2}$ where
$\tilde{F}_1 (\pi)$ is the pionic contribution to the $F_1$
Landau-Migdal parameter which has the value $\frac 13 \tilde{F}_1
(\pi)=-0.153$ at nuclear matter density. With $\Phi
(n_0)=\frac{m_V^\star (n_0)}{m_V}\approx 0.78$, one obtains
$g_A^\star/g_A\approx 0.80$ or $g_A^\star\approx 1.0$. It must be
mentioned here that how these descriptions are related is not
understood yet and remains an open theoretical issue. Now in
finite nuclei the $g_A^\star$ is found almost universally to be
unity~\cite{kooninetal}. Thus,
 \be
m_N^\star/m_N\approx
(\frac{1}{g_A})^{1/2}\frac{m_\rho^\star}{m_\rho}\approx 0.7 \ \ \
{\rm at}\ \ n=n_0.
 \ee
We suggest that this is equivalent to the nucleon mass scaling
needed by Friedman. In our formulation, the phenomenological
approach of Friedman corresponds to the leading-order treatment of
our effective theory with the vector
manifestation~\cite{HY:PR,BR:PR01} for which the only scaling
parameter is the pion decay constant with the pion mass
unscaled~\footnote{In HLS/VM with the explicit vector degrees of
freedom, the scaling parameters are the gauge coupling $g^*$, the
vector-meson mass $m_V^*$, the pion decay constant $F_\pi^*$ and
the coefficient $a$. However in medium $a^*\approx 1$, $g^*\approx
g$, so $m_V*$ goes as $F_\pi^*$ through HLS/VM relations. We are
left with only one scaling factor as in the case where the vectors
are integrated out.}.

Note that we arrived at (\ref{prediction}) without making direct
reference to the quark condensate. It was based on the
``many-body" relation presumably valid up to and near nuclear
matter density and not beyond the Fermi-liquid fixed point.
Relying on (\ref{condensate}), one might naively conclude that the
result signals a reduction of the quark condensate with density
and hence might be taken as the signal that chiral symmetry is
being restored. This inference is valid however only in the tree
approximation used here and does not apply when one works at
higher (loop) orders. The bare pion decay constant fixed at the
matching scale does not necessarily follow the quark condensate
near chiral restoration as density is increased unless one takes
into account the parametric dependence that follows the
renormalization group flow. This is so even though the physical
pion decay constant vanishes at the chiral transition in the
chiral limit. With the effective Lagrangian that most of the
people in the field use, namely that in which the parametric
dependence is absent, the behavior of the ratio as a function of
density has little to do with chiral restoration, a point which
seems to be often overlooked.

\section{The Vector Manifestation and the Photon Coupling to
the Nucleon}\label{photon} \itt It turns out that in HLS/VM, the
photon coupling to hadrons in medium can be quite different from
that in free-space. For instance it has been shown by Harada and
Sasaki~\cite{HS:VD} that the vector dominance in the photon
coupling is strongly violated near the VM fixed point and hence
near the chiral transition point. While experiments at Jefferson
Lab probe densities near that of nuclear matter and hence rather
far from the VM fixed point, the presence of nucleon in the medium
is expected to drive the parameter $a$ in HLS Lagrangian toward 1
from its value $a=2$ at the vector-dominance regime~\footnote{The
deviation from vector dominance in the photon coupling to the
baryons has been known since some time. In fact, the chiral bag
model~\cite{CND,RGB} provided a natural mechanism for such a
deviation.}. This would have a strong ramification on properties
of in-medium nucleon form factors. In this section, we briefly
discuss what can be expected.

In HLS, the photon couples to the degrees of freedom involved in
the theory as
 \be
\delta \calL= -2eag F_\pi^2 A^\mu {\tr}[\rho_\mu Q]
+2ie\left(1-\frac a2\right) A^\mu{\tr} [J_\mu Q]\label{coupling}
 \ee
where $A^\mu$ is the photon field, $Q$ is the electric charge
matrix $Q=\frac 13\ {\rm diag}\ (2 -1 -1)$ and $J_\mu$ is the
vector current made up of the chiral field $\xi$. In HLS theory,
baryons could be thought of arising as skyrmions in which case
$J_\mu$ can be identified as the skyrmion current. Alternatively
one could introduce ``bare" baryon or quasiquark field into the
HLS scheme in which case, one can think of $J_\mu$ as a ``bare"
baryon current. The usual vector dominance (VD) picture
corresponds to taking $a=2$ in which case there will be no direct
coupling of the photon to the baryon just as there is none in the
case of the pion (for which the second term of (\ref{coupling})
would be of the form $J_\mu=[\del_\mu \pi,\pi]$). It turns
out~\cite{HY:VDM} that the vector dominance (VD) at $a=2$ is on an
unstable trajectory of renormalization-group flow of HLS theory
with no connection to the trajectory that leads to the VM and that
the fact that in nature the VDM seems to work in matter-free space
and at $N_F=3$ is merely an ``accident." In fact, in the presence
of matter (temperature or density), the flow consistent with QCD
is on the trajectory that leads to the VM fixed point $a=1$. Even
in the absence of matter, $a\approx 1$ seems close to nature
although the VDM that works for the EM pionic form factor
corresponds closer to $a=2$. See \cite{HTY:pion}.

The vector dominance picture is not a good one for the photon
coupling to the nucleon, so $a=2$ must hold poorly when nucleons
are involved. Indeed, there is an indication that the photon
coupling to a single nucleon is already near this fixed point.
Historically a picture closely resembling this one was adopted by
Iachello, Jackson and Lande in 1973~\cite{iachelloetal}. There the
authors assume that the $\gamma$-ray couples to nucleon more or
less equally (at low momentum transfer) through the vector meson
and directly to a compact core. The presence of the small core in
the proton, of $0.2\sim 0.3$ fm is indicated also in the proton
structure function in (deep) inelastic
scattering~\cite{structuref}. In fact this is the 50/50 picture
that arises at the ``magic angle" in the chiral bag
model~\cite{RGB,CND} used by Soyeur, Brown and Rho~\cite{SBR1,BRS}
in analyzing nuclear form factors. In medium, we predict that the
electromagentic form factor will have the form for $a=1$,
 \be
\frac e2 \frac{1}{1-q^2/{m_V^*}^2} + \frac e2 H (q^2)
 \ee
where $H$ is a slowly varying function of $q^2$ (with $q$ being
the four-momentum transfer) with $H(0)=1$ and $m_V^*$ is the
parametric mass that enters at finite density. The photon point
$q^2=0$ gives the correct charge. The $\gamma$-ray will couple to
the dileptons in this half-way manner. Note that the dileptons
discussed below will experience the same propagator suppression,
namely, the correction factor
 \be
F\approx \frac{1+1/(1+Q^2/{m_V^*}^2)}{1+1/(1+Q^2/{m_V}^2)}
 \ee
with $Q\equiv |\vec{q}|$ but $Q$ is generally small, the dileptons
being nearly back to back, so this is probably unimportant. Thus
though vector dominance is violated (expected in the nucleon
sector from other considerations than that of the VM), it is an
adequate approximation in the dilepton calculation as we will
discuss below.
\section{``Sobar" Configurations and CERES Dileptons}\label{sobar}
 \itt
 There have been much discussions on the possible evidence for
changes in  hadron properties in the CERES dilepton
data~\cite{rapp}. While the processes discussed above involve
transition matrix elements with specific kinematics, the dilepton
experiments measure spectral functions or more generally
correlation functions averaged over density and temperature. For
this purpose, we need to look at the spectral distribution in the
hot and/or dense environment. In terms of the framework based on
HLS/VM theory we are adopting in this paper, this means that we
need to incorporate consistently into vector-vector correlation
functions both quantum fluctuations with the parametric masses and
coupling constants and thermal loop and/or dense loop effects
generated in the renormalization-group flow. Now as pointed out in
\cite{BLRRW,KRBR}, this means, for the dilepton processes,
considering both BR scaling figuring in density-dependent
parameters that results from the CSD and the mixing to the
``sobar" configuration $N^* N^{-1}$ computed as ``fluctuations" on
top of the soliton configuration. This requires that the
double-decimation be consistently implemented. We shall refer to
this procedure as ``BR/RW fusion." In the Rapp-Wambach approach
(abbreviated as R/W)~\cite{rapp}, the fusion has not been
implemented: There, the second decimation is replaced by
configuration mixing in lowest-order perturbation theory while the
first decimation CSD is left out.

In several low-energy phenomena the R/W \rhosobar\ provides most
of the low-lying ``$\rho$" strength as we shall outline, B/R
coming in as the $intrinisc$ effect to somewhat increase the
effect. This is because the \rhosobar\ has $\sim$ 20\% of the
$\rho$-strength, at a low energy of 580 MeV. BR scaling can only
move the remaining 80\% at the parametric $\rho$-mass to this
energy at densities $> n_0$ (nuclear matter density), and such
high densities have not been investigated experimentally. The
\rhosobar\ pushes the elementary $\rho$ up in energy, the states
repelling each other; thus, in the region of the elementary $\rho$
the \rhosobar\ and the $\rho$ ``defuse." We shall see in Section
\ref{reginarhosobar} and Section \ref{RHIC} that there is good
empirical evidence for this, establishing that both R/W and B/R
are present as required.

\subsection{The role of the ``$\rho$-sobar" in the $^3$He$
(\gamma,\rho^0)ppn$ reaction}\label{reginarhosobar}
 \itt Before we treat the CERES dilepton production which will
require the fusion of the ``$\rho$-sobar" (or R/W) and BR scaling
(i.e., B/R), we consider the process where the $\rho$-sobar plays
the primary role. Two recent papers by Lolos et
al~\cite{lolos1,lolos2} have discovered the fact that the $[N^*
(1520)N^{-1}]$ excitation decays $\sim$ 20\% of the time into a
$\rho$-meson, in agreement with the 15 - 30 \% listed in the
Particle Data Book and the $\sim 20$ \% found by Langg\"artner et
al~\cite{lang}. We call the $[N^* (1520)N^{-1}]^{1^-, I=1}$ the
``$\rho$-sobar" because when measured in finite nuclei, rather
than when produced on a single proton by, e.g., $\pi +p$
interactions, it takes on a collective character with increased
$\rho$-meson content due to the admixture of the elementary $\rho$
with density~\cite{BLRRW}.

A clear evidence has been found that the $\rho$ production in the
deuteron is dominated by the $N^* (1520)$, an element of the
$\rho$-sobar. Of course the nucleon density of the deuteron is so
low that one could hardly expect any appreciable medium dependence
from this nucleus. Greater $\rho$ production should be seen in
$^3$He, although the present experimental accuracy does not seem
to be sufficient to show this. Here we give theoretical estimates
for how much greater it should be with Rapp/Wambach alone and with
Brown/Rho fused with Rapp/Wambach. These estimates can be easily
extended to heavier nuclei.

From the $^3$He wave functions of Papandreou et al~\cite{pap} we
easily see that the average density in $^3$He is half that in
nuclear matter. Although these authors could explain the
experimental results with BR scaling alone, with our analysis we
find that they had to decrease the $\rho$ mass twice too much by
their mean field. As we shall develop, R/W which they also
considered gives most of the experimental effects.

From \cite{BLRRW} -- that we shall refer to as BLRRW -- we see
that the mixture of the elementary $\rho$ into the $\rho$-sobar,
which goes as the square of the matrix element divided by the
energy difference increases linearly with the density $n$. This is
because the \rhosobar\ is a collective state, a linear combination
of the excitations of all nucleons in the nucleus up into the
\nstar. The amount of $\rho$ admixed into the \rhosobar\ is the
same as the amount of \rhosobar\ admixed into the $\rho$. This is
displayed, as function of density, in Fig.1 of Kim, Rapp, Brown
and Rho~\cite{KRBR} in which it is shown that the sobar
configurations can conveniently be incorporated into a massive
Yang-Mills theory. For $n=n_0/2$, the $Z$ factor comes out to be
$Z_\rho=0.15$ for R/W; fusing with B/R increases it to 0.23. This
gives the increased $\rho$ content of the $in$-$medium$ \rhosobar,
which then increases from 20\% found by Langg\"artner et al (15 -
25\% in the Particle Data Booklet) to 40\% with the fused R/W and
B/R. If the same were to be obtained by R/W alone, the density
would have to be increased by $\sim$ 50\%.

The recent unpublished calculation on the Regina
$^3$He$(\gamma,\rho^0)ppn$ reaction by Rapp
show that compared with experiment~\cite{haber},
the R/W results are spread widely about
the unperturbed zero-density \rhosobar\ energy of 580 MeV,
somewhat more upwards than downwards, by the large imaginary part
which we estimate to be $\gsim$ 200 MeV, 150 MeV zero-density
width of the $\rho$ plus some in-medium width. The \rhosobar\
width increases with energy spreading the strength more upwards.
On the whole the R/W fit is good, except that too much $\rho$
strength is predicted in the region up to the elementary $\rho$ at
770 MeV. This will be improved by the fusing with B/R, the latter
lowering the parametric $\rho$ mass in the Lagrangian by 77 MeV.
The definite need for the $fusing$ of R/W and B/R is seen,
therefore, most clearly in the region of the elementary $\rho$
where the push upwards by R/W must be compensated for by the
downward shift from B/R. Here R/W and B/R ``defuse" for the
elementary $\rho$. We return to this point in the next section on
RHIC physics.

Asked by the experimentalists, Rapp extended his calculations to
higher densities than present in $^3$He and $^{12}$C
in order to increase the medium effect. As noted, an $\sim$ 50\%
increase in R/W in the \rhosobar\ region is effected by the fusion
with B/R, which we believe to explain the greater collectivity
apparently seen.

While the fusion of B/R with R/W does give a good description of
the observed phenomena, the accuracy in measurements of the
Langg\"artner et al and Regina-Tokyo groups is not great enough
yet to show the necessary increase in $\rho^0$ production cross
section per nucleon in going from proton to $^3$He targets from
medium effect. In Huber et al~\cite{lolos2}, the cross section for
photo-production of the $\rho$ of $10.4\pm 2.5\ \mu b$ on the
three $^3$He nucleons is found for the 620 - 700 MeV range of
photon energy ($m_\rho^*\sim 500$ MeV) whereas Langg\"artner et
find $\sim 3.5\ \mu b$/nucleon for $\rho^0$ production. Possible
uncertainties in the latter cross section are not given, but if
they are comparable in magnitude to the Regina uncertainties, even
with the upper limit of Regina and the lower limit of
Langg\"artner et al we could not achieve our estimated medium
effect.

In the $^3$He $(\gamma,\rho^0)ppn$ the low-mass $\rho$ strength is
mostly explained by R/W, the B/R coming in to slightly enhance the
low-mass strength, but chiefly to lower the high-mass $\rho$
strength, which in R/W alone fails to explain experiment. Thus the
higher energy region in which R/W and B/R defuse shows
definitively that both effects are present. (See the next
section.)
\subsection{The CERES: Fusion of B/R and R/W}
 \itt
We now go on to the fusion of the two effects described above in
the dilepton production at CERN by the CERES collaboration in
which R/W and B/R play about equal roles, the present error bar
and binning in the experiment being too large to pin down their
separate roles.

Ralf Rapp (private communication) has carried out calculations
fusing B/R and R/W, using vector dominance. 
Although it has been shown in
HLS/VM~\cite{VDviol} that vector dominance is violated at high
temperature, as noted above, this should not affect the numerical
results.
%
%
From the Rapp results, one can say with some confidence that the
fit is improved with the fusion of B/R and R/W. However due to the
error bars, one cannot say firmly that the fusion is $required$.
As it stands, both R/W
and B/R
fit the data just as well within the error bars. It is clearly
difficult to differentiate their separate roles by experiment in
the low-energy regime. However we shall see from the next section
that this separation is straightforward in the high-energy region
where they defuse.
\section{RHIC Resonances}\label{RHIC}
\itt It should be clear in our discussion of the dileptons given
above, also $\rho$-mesons produced by $\gamma$-rays on $^3$He and
other nuclei, that it is intricate to separate the role of BR
scaling from that of Rapp-Wambach. It is therefore significant and
exciting that a direct measurement of the $\rho$-meson
mass~\footnote{The ``mass" involved here is the pole mass. However
the difference between the parametric and pole is of higher order
in the power counting and can be taken to be negligible for the
discussion.} $m_\rho^*$ could be made in a pristine atmosphere
where temperature effects are small and where the density can be
well reconstructed.

\begin{figure}[htb]
\vskip 1cm
 \centerline{\epsfig{file=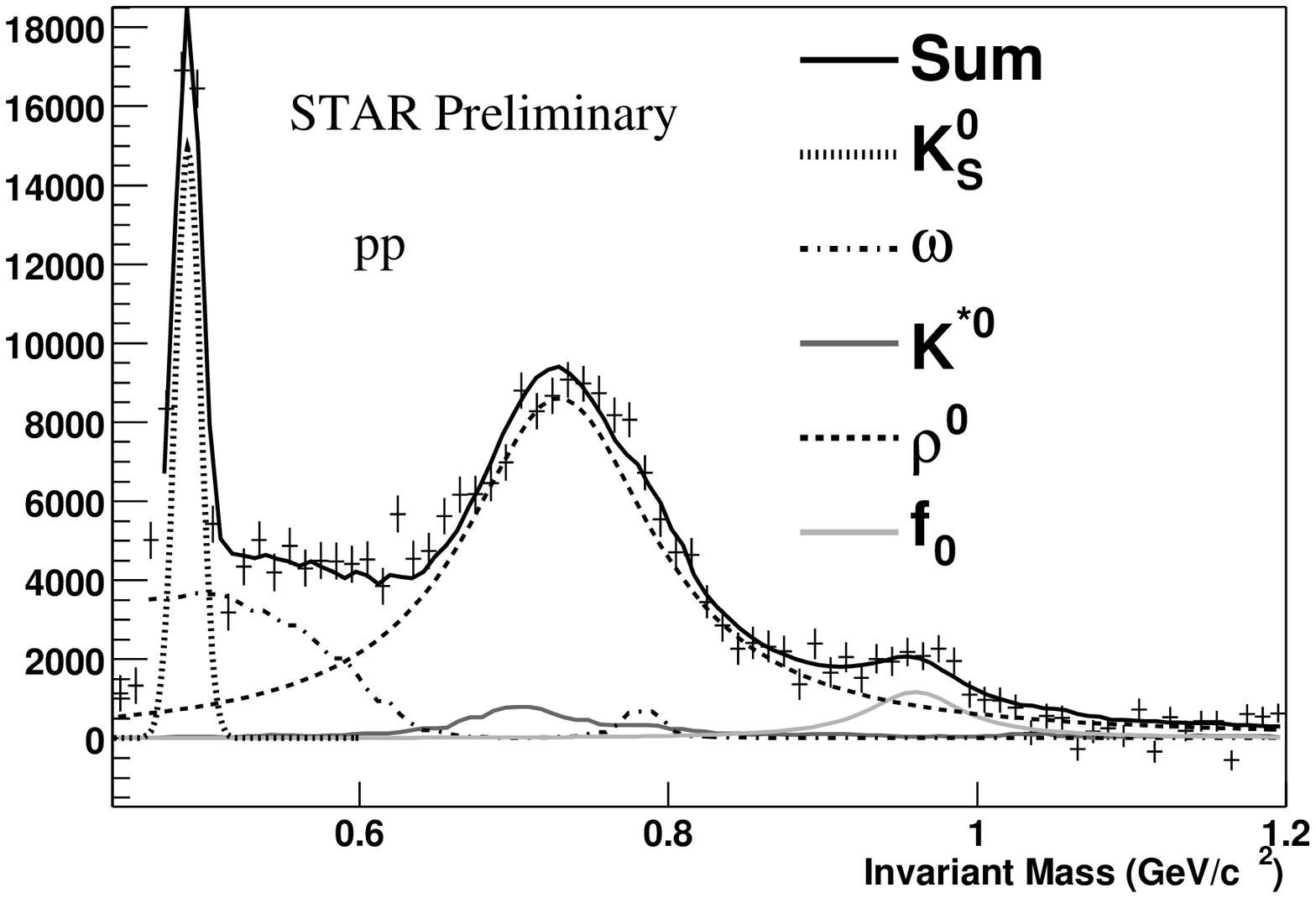,width=7.5cm,angle=0}
 \epsfig{file=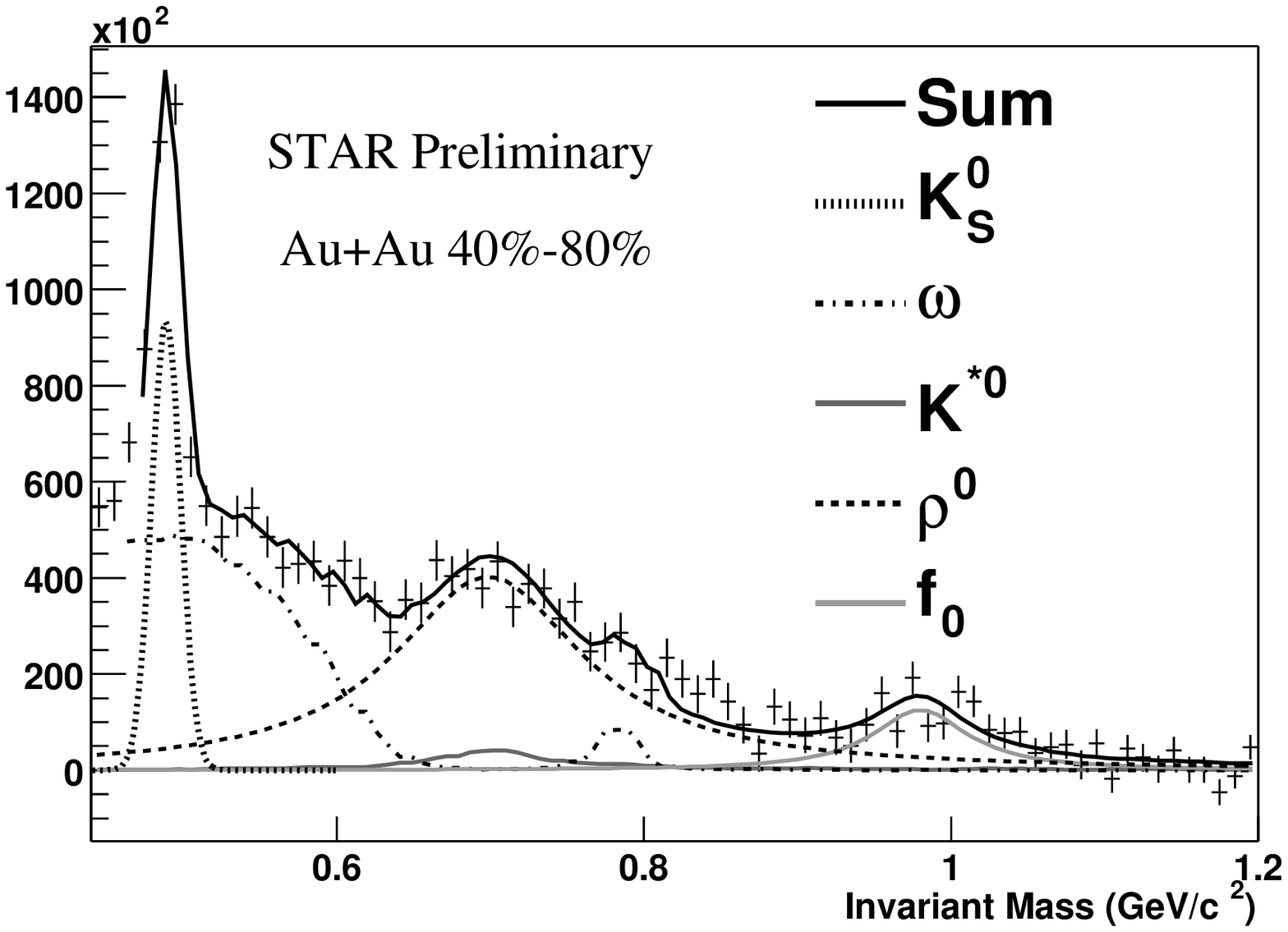,width=7.5cm,angle=0}}
 \vskip 0cm
\caption{The invariant mass distributions in pp and mid-central
AuAu of the $\pi^+\pi^-$ system, with a transverse momentum cut
$0.2 <p_t <0.9\ {\rm GeV}$. From Fachini~\cite{fachini}.
Contributions from specific resonances are indicated by different
lines. }\label{XX}
\end{figure}

We show in Fig.\ref{XX} the STAR/RHIC results~\cite{fachini}. The
(preliminary) fits gave
 \be
m_\rho&=& 0.698\pm 0.013\ \ {\rm GeV} \ \ {\rm for}\ \
Au-Au,\nonumber\\
&=& 0.729\pm 0.006\ \ {\rm GeV} \ \ {\rm for}\ \ pp.
 \ee

The collisions were measured at $\sqrt{s}=200$ GeV. There is a
long history of the detailed $\rho$ spectral shape. The difference
between its appearance in elementary reactions, $e^+e^-\rightarrow
\pi^+\pi^-$ or $\tau\rightarrow \nu_\tau\pi^+\pi^-$ and
hadroproduction reaction is well known. The latest Review of
Particle Physics~\cite{reviewparticle} averages the $\rho$ mass
for these two sets of experiments separately, with a clear
systematic difference of the order of 10 MeV between them:
 \be
m_\rho^{leptoproduction}&=& 775.9\pm 0.5 \ \ {\rm MeV},\nonumber\\
m_\rho^{hadroproduction}&=& 766.9\pm 0.5 \ \ {\rm MeV}.
 \ee

It was noticed back in the 1960's by Hagedorn and others that the
particle composition in $pp$ can be well reproduced by statistical
models, see, e.g., ref.\cite{becattini}. The fitted temperature is
about the same as the chemical freezeout temperature of $\sim 165$
MeV found at RHIC. In some sense the $\rho$-meson must be born in
a heat bath, with Boltzmann factors which cut off the high energy
end of the rather broad $\rho$. Some discussion of this can be
found in Shuryak and Brown~\cite{shuryakbrown} and in Kolb and
Prakash~\cite{kolb}. We expect that many papers will be written on
this subject, which is not within the scope of this paper. We can
only assume that perhaps half or less~\cite{kolb} of the drop in
the $\rho$-mass found in the Au-Au experiments comes from the
heat-bath-related effects, and that these are about the same as in
the pp experiment.

Next we discuss shifts in the $\rho$-mass from forward scattering
such as $\rho+\pi\rightarrow a_1\rightarrow \rho+\pi$. There are
many of these which enter as principal values as discussed in
\cite{eletskyetal,rappgale} as well as in \cite{shuryakbrown}. It
seems difficult to get more than a few MeV out of these, the sign
probably attractive. We note that a substantial part of the upward
push in these resonance calculations comes from the \rhosobar\
($N^* (1520)N^{-1}$) which gives the Rapp-Wambach effects. Thus
the RW and BR ``defuse" for the elementary $\rho$. In other words,
the BR scaling must substantially overcome the Rapp-Wambach
configuration mixing in order to produce a substantial downward
shift. In fact the BR scaling shift will be, from section
\ref{d1-scaling} (see \cite{BR:qm02}),
 \be
m_\rho^*\simeq m_\rho(1+0.28n/n_0)^{-1}
 \ee
where $n$ is the total baryon density in nonstrange
particles~\footnote{In Shuryak and Brown~\cite{shuryakbrown}, the
mass shift was associated, in the language of linear sigma model,
with the ``amplitude-field" fluctuation of the chiral field -- or
radius field $R$ --, not with the ``phase-field" or pionic
fluctuation. It is perhaps worthwhile to point out that this way
of looking at the mass shift is totally equivalent to the
nonlinear sigma model approach of Brown and Rho~\cite{BR91} where
the ``dilaton" field of trace anomaly $\chi$ plays the role of the
amplitude field in \cite{shuryakbrown}. This point has been
further clarified in \cite{LPRVchi}. In fact, it is shown that the
approach of \cite{BR91} is closer to the modern treatment based on
HLS/VM theory of Harada and Yamawaki. In the language of HLS/VM,
it is the intrinsic parametric dependence on density resulting
from matching to QCD that plays an essential role in the mass
scaling. The pionic fluctuation in the theory is subject, of
course, to low-energy theorems of chiral symmetry.}. This comes
about because for low densities $m_\rho^*$ scales as $f_\pi^*$
while the quark mass and the pion mass do not scale.

As a final approximation in determining the density at which the
$\rho$-meson decays we might consider thermal (kinetic) freezeout,
since the two $\pi$'s must come unscattered to the detector. The
kinetic freezeout at RHIC can be obtained from a hydro-based fit
to the data~\footnote{P. Kolb, private communication.}
 \be
T_k\approx 100\ {\rm MeV}, \ \ \mu_\pi\approx 81\ {\rm MeV},\ \
\mu_N\approx 380\ {\rm MeV}, \ \ \mu_K\approx 167\ {\rm MeV}
 \ee
which translate into the pion density $n_\pi\approx 0.06\ {\rm
fm}^{-3}$ and
 \be
n_{N+\bar{N}}\approx 0.0075\ {\rm fm}^{-3},
 \ee
or only 1/20 of nuclear matter density.

\begin{figure}
\vskip 0cm
 \centerline{\epsfig{file=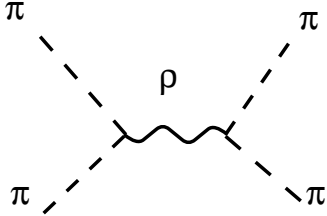,width=15cm,angle=0}}
 \vskip -7cm
 \caption{The product of probability of $\rho$ formation and decay
into pions that can get out without being scattered must be
optimal. }\label{XXX}
\end{figure}

To obtain a better approximation we must optimize the product of
the $\rho$ resonance formation and its decay, as shown in
Fig.\ref{XXX}. Ignoring the lifetime of the $\rho$ in comparison
with the other times -- which underestimates the density at which
the $\rho$ is formed since all the $\rho$'s will follow the
fireball expansion -- the parameters for resonance formation and
decay are found by Shuryak and Brown~\cite{shuryakbrown} to be
 \be
T_k\approx 120\ {\rm MeV}, \ \ \mu_\pi\approx 62\ {\rm MeV},\ \
\mu_N\approx 270\ {\rm MeV}, \ \ \mu_K\approx 115\ {\rm MeV}
 \ee
with densities about 1.4 times higher than at kinetic freezeout.
Taking into account that the density of excited baryons is double
that of $N+\bar{N}$, and the above conditions, we find the total
density in nonstrange baryons to be 0.21$n_0$.

We note that in BR scaling, basically the total density in
nonstrange quarks and antiquarks should be used, whereas in RW
configuration mixing the assumption of these authors is that the
resonances with $\rho$ quantum numbers built upon excited baryons
will be only $\sim 50$ \% as efficient in configuration mixing as
those built upon the nucleon. Thus, at a total baryon density with
2/3 made up of excited baryons, they would add the 1/3 in
$N+\bar{N}$. In this way BR becomes somewhat larger than (and of
opposite sign to) RW under RHIC condition.

At the resulting density $n\approx 0.21 n_0$, and with the
$\rho$-mass dropping $\sim$ 22 \% at $n_0$, we find a decrease of
$\sim$ 43 MeV in $m_\rho^*$. However, the $\rho$-meson themselves
-- as well as the $\omega$'s , to a lesser extent -- are also a
source of scalar density, i.e., they are composed of constituent
quarks~\footnote{Since pion is not a source of scalar density,
although composed of constituent quarks, cancellations must insure
that its mass is not changed; these in turn insure that the pions
are removed as source of scalar density.}.

Considering the $\rho$'s, we note that equilibrium through
$\rho\leftrightarrow 2\pi$ continues basically down to thermal
freezeout, and with $\mu_\pi\approx 62$ MeV, the pion fugacity
will increase the number of $\pi$'s by $e^{\mu_\pi/T}$ and the
number of $\rho$'s by the square of this factor. With the greater
multiplicity, but smaller Boltzmann factor than the $\pi$'s, we
find $n_\rho/n_\pi\sim 0.15$, with $n_\pi\sim 0.08/{\rm fm}^3$, or
an $n_\rho\sim 0.013/{\rm fm}^3$ which considering that the $\rho$
presents 2/3 of the scalar source as a nucleon, gives the
additional $\sim 10$ MeV to the downward shift of the
$\rho$-meson. Thus our total shift from BR scaling is $\sim 53$
MeV, roughly 80\% of the total observed downward
shift~\footnote{We should note that the two scalings (\ref{br1})
and (\ref{Phi}) which coincide for $n=0$ and $n=n_0$ give results
that differ by $\sim 13$ MeV in the shift at $n\approx 0.21 n_0$
relevant to the process in question. The shift given here is
somewhat larger than that quoted by Shuryak and
Brown~\cite{shuryakbrown}. This difference cannot be taken
seriously as we really do not know precisely how the scaling
interpolates between $n=0$ and $n=n_0$.}. This seems to account
for most of the shift left over when the temperature-driven shift
is subtracted away.

The 150 MeV width of the $\rho$ is unchanged. Since the decay is
p-wave, the width should decrease with the cube of the momenta of
the pions into which the $\rho$ decays. This would cause an $\sim$
30\% decrease in width. The fact that the width does not change
means that compensating increases must be furnished by the
resonances, both those above and below the $\rho$ in mass
contributing with equal sign to the width. From the upper part of
Fig.1 of Rapp and Gale~\cite{rappgale}, one can see that their
resonances contribute $\sim 50$ MeV to the total width for mass
$M=700$ MeV at $T=150$ MeV whereas in \cite{eletskyetal}, the
increase is somewhat larger. In any case, the STAR results give a
nice check that the increase in width of  $\sim 50$ MeV (collision
broadening) is cancelled by the decrease in width from lower
penetrability.

The RHIC work has the great advantage over the dileptons that the
energy resolution separates off the upper region in which BR and
RW defuse, showing that BR must first cancel RW  and then add some
net downward shift to the $\rho$-mass, the general size predicted
by BR scaling fitting in nicely with the experimental results.

\section{Effective Forces in Nuclei}\label{effectiveforce}
 \itt
The $V_{low-k} (r)$ has supplanted the Kuo-Brown
G-matrix~\cite{kuo-brown} as effective interaction to be used in
nuclear structure calculations. Integrating-out of the high
frequency parts of the two-body interaction supplants all of the
paraphernalia of off-shell particle energies, etc~\cite{BKCCI}.
Here we discuss how the CSD (chiral symmetry decimation) may
change some of the results of the FLD (Fermi liquid decimation).

The most extensive, but still only partial, investigation of the
problem we discuss here was carried out by Hosaka and
Toki~\cite{hosaka} for the $2s,1d$-shell matrix elements (with
$^{16}$O as closed shell). In this region of nuclei, up to
$^{40}$Ca, empirical two-body matrix elements which fit well the
experimental spectra have been determined. The most definite and
important result of Hosaka and Toki is that ``the central G-matrix
elements are already well reproduced by using the free-space
parameters and that they are extremely sensitive to the masses of
$\sigma$ and $\omega$ mesons. This indeed provides a strong
constraint from the phenomenological side on the way the meson
masses, particularly $m_\sigma^\star$ and $m_\omega^\star$ should
scale in medium. In fact, it turned out that they have to be
correlated such that $m_\sigma^\star/m_\sigma\sim
m_\omega^\star/m_\omega$ as long as the coupling constants are
kept unchanged." Note that the latter conclusion is consistent
with what we have found in \cite{BR:qm02} and with the type-II
analysis of nuclear matter by Song~\cite{song}. As for the former,
it can be explained by the argument presented in the Introduction,
namely the effect of the strong $\rho$-meson tensor coupling which
more than cancels the strong pionic tensor coupling at short
distances and keeps the nucleon-nucleon potential small in
magnitude (except for the strong short-range repulsion which is
counterbalanced by the short-range attraction and integrated out
to give $V_{low-k} (r)$). Thus off-shell effects are small.

``Encouraged by this fact" (i.e., that with BR scaling the central
G-matrix elements were well reproduced), Hosaka and Toki ``have
calculated the G-matrix elements using the various meson-nucleon
masses scaled nearly the same way while keeping the pion strength
unchanged. This time our interest is concerned with the LS and
tensor matrix elements, since they are generally in poor agreement
with nature if the free-space masses are employed." By using the
scaled masses, however, they ``could not see significant
improvements in the comparison of the calculated and empirical
matrix elements." They find that generally the LS matrix elements
are enhanced by the increase of the $\sigma$, $\omega$ and $\rho$
contributions and the tensor matrix elements are suppressed by the
increase in the repulsive components due to $\rho$ exchange.

The little improvement from the dropping masses in the spin-orbit
channel is disappointing, given the excellent agreement found by
Tjon and Wallace with or without dropping masses in Fig.\ref{fig1}
and \ref{fig2}. However, the rapid oscillations characteristic of
the Dirac phenomenology, as compared with the nonrelativistic
equations, come about from the interference in contributions from
different densities in the nucleus. Such an effect would be
missing in taking the scaled masses to be constant, independent of
local density, as Hosaka and Toki did.

It was actually the decrease in the in-medium tensor interaction
found in the 1989 and 1990 Brown-Rho papers which provided an
empirical guide for the dropping $m_\rho^\star$.

The strong spin-orbit interaction, which is the basis for the
shell model, was obtained in the Walecka-type relativistic mean
field calculations, a factor $\sim 2$ at $n=n_0$ greater than
found in nonrelativistic Brueckner calculations. This spin-orbit
interaction is very important in the single-particle energies,
giving the splitting between the spin-orbit partners in the shell
model. The Walecka theory gets sufficient spin-orbit splitting by
using nucleon effective mass $m_N^\star\sim 0.6 m_N$, somewhat
smaller than what we find, $m_N^\star\approx 0.7 m_N$. We actually
get $m_N^\star$ this low by assuming $g_A^\star\approx 1$ in
medium. Since one does not expect that the Gamow-Teller coupling
constant (globally) goes below 1, this is the lowest we can
expect.

The spin-orbit interactions due to scalar and vector exchanges
that Hosaka and Toki took are of the form
 \be
V^{LS}_S&=&-\frac{g_S^2}{4\pi} \frac{m_S}{2} (\frac{m_S}{m_N})^2
Z_1(m_Sr),\nonumber\\
V^{LS}_V &=& -\frac{g_S^2}{4\pi} m_V (\frac{3}{2}
+\frac{2f_V}{g_V})(\frac{m_V}{m_N})^2 Z_1(m_Vr),
 \ee
where
 \be
Z_1 (x)=(1+1/x) \frac{e^{-x}}{x}.
 \ee
Note that the volume integrals of $V^{LS}_{S,V}$ are independent
of $m_S$ and $m_V$ respectively, the masses going into rescaling
the $r^2dr$ factor in the integral.~\footnote{This is also true
for the central potentials.} In mean field, energies etc. depend
chiefly on the volume integrals. However, an additional $m_N^{-2}$
is present in the spin-orbit term (also in the tensor
interaction). Taking $m_N\rightarrow m_N^\star$ would strongly
enhance both of these interactions, say, by a factor of $\sim 2$
at $n=n_0$. Of course, the average density in the $s,d$-shell lies
below $n_0$.

As noted, the need for nucleon effective masses was clear in
Walecka theory and these effective masses fixed up the spin-orbit
interaction. If $m_N^\star$ scaled like $m_S^\star$ or
$m_V^\star$, the scale invariance noted in Section
\ref{diracpheno} would still hold, being violated only by the
constancy of the pion mass. But because of the ``loop correction"
 \be
\frac{m_N^\star}{m_N}\approx\sqrt{\frac{g_A^\star}{g_A}}\frac
{f_\pi^\star}{f_\pi}
 \ee
and since $g_A^\star$ tends to unity precociously in nuclei, the
$m_N^\star$ scales more rapidly than $f_\pi^\star$ just up to
nuclear matter density. So the $\sqrt{\frac{g_A^\star}{g_A}}$
factor breaks the scale invariance, but judging from the
Tjon-Wallace results, not in a serious way.

We therefore conclude that the tensor force is the only component
in nuclear force to be substantially changed by BR scaling,
bearing out the 1990 work~\cite{BR-stiff90}. Along the same lines
and even earlier~\cite{brownetal} evidence had been found for an
in-medium $\rho$ with a factor of 1.5 times that in the free
G-matrix~\cite{brown2}. This came from a study of 447 sd-shell
binding energy data. Assuming the strength of the $\rho$-exchange
potential be proportional to strength times (range)$^2$, this
would correspond to $m_\rho^*\sim 0.82 m_\rho$ if the strength is
kept constant and the range changed. In 2s, 1d-shell nuclei the
average density is well below $n_0$, so one would not expect the
$m_\rho^*$ to drop quite so much. In any case Brown et
al~\cite{brownetal} found this change to improve agreement with
data.

It appears difficult to find ``smoking guns" in {\it the structure
of nuclei} for 20\% changes in hadron masses in going from $n=0$
to $n=n_0$ implied by BR scaling. In the case of $\omega$- and
$\sigma$-exchange we saw from the Hosaka-Toki work that the
effects cancel each other. However, the $\rho$ meson has no
low-mass isoscalar partner, and one might look elsewhere than in
the tensor interaction for effects, especially in the symmetry
energy, which is so important in astrophysical applications.

In mean field theory the effect of the vector-coupled $\rho$ on
the symmetry energy at nuclear matter density would be an increase
by $(m_\rho/m_\rho^*)^2\approx (0.8)^{-2}=1.56$. However,
modifications in the $\rho$-mass also affect the second-order
tensor interaction, which, as we remarked in the Introduction,
gives the main difference between $^3S_0$ and $^1S_0$ states.

It was shown by Kuo and Brown~\cite{KB65} that the closure works
well to approximate the second-order tensor interaction; i.e.,
 \be
V_T \frac{Q}{E} V_T\approx \frac{V_T^2}{E_{eff}}
 \ee
with $E_{eff}\approx 250$ MeV~\footnote{Note that $E_{eff} > 175$
MeV, the c.m energy corresponding $p=2.1$ fm which is the cutoff
that figures for $V_{low-k}$. In EFT, this means that the second
order tensor effect should reside predominantly in the local
counter term.}. Now
 \be
V_T^2\approx -\frac{1}{250{\rm MeV}} (3-\tau_1\cdot\tau_2)(6+2{\bf
\sigma}_1\cdot\boldmath{\sigma}_2- 2 S_{12}) V_{T(\pi+\rho)}
(r)\label{VT2}
 \ee
where the final factor is the square of the radial part of the
tensor term. In getting (\ref{VT2}), we have used the identity
${\bf {\sigma}}\cdot A \sigma\cdot B=A\cdot B+i\sigma\cdot
[A\times B]$ and omitted the term linear in $\sigma$ which will
vanish when averaged over angle. We see that in the triplet state
the second-order tensor gives the contribution
 \be
-\frac{24}{250{\rm MeV}} V^2_{T(\pi+\rho)} (r)
 \ee
which explains why the triplet state is bound, and the singlet
not. However this is also the contribution to the symmetry energy
 \be
\delta V_{symm}=\frac{6\tau_1\cdot\tau_2}{250{\rm MeV}}
V_{T(\pi+\rho)}^2 (r),
 \ee
which comes to $\sim 1/4$ of the central contribution.
Relativistic Brueckner-Hartree-Fock calculations with the strong
$\rho$-coupling of the Bonn potential~\cite{mutheretal} show the
$\rho$-contribution to $\delta V_{symm}$ just cancels that from
the vector coupling of the $\rho$, leaving the pion exchange
(mostly in second order) on the source of symmetry energy. Thus we
seem to be foiled, again, in nuclear structure physics a strong
``smoking gun" for the BR scaling (or parametric dependence on
density) of the $\rho$ mass.

The conclusion is that in all nuclear structure observables so far
probed, the parametric density dependence symptomatic of BR
scaling is masked in an intricate way. This strongly suggests a
``hidden symmetry" which seems to induce almost exact cancellation
of possible smoking-gun signals, a phenomenon that is very much
reminiscent of the Cheshire-Cat phenomenon expounded in
~\cite{CND}.
\section{Conclusion: Return of the Cheshire Cat}\label{concl}
 \itt
The main conclusion we have arrived at in this paper is that up to
date, no ``smoking gun" signals have been observed in nuclear
structure physics for BR scaling indicative of the spontaneous
breaking or restoration of chiral symmetry, a basic element of
QCD. This outcome, although somewhat disappointing and perhaps
unappealing, is however very much in line with what we have been
arguing since some years, namely, that low-energy strong
interactions are strongly governed by the (approximate)
Cheshire-Cat Principle (CCP). Here we would like to enumerate a
few cases where the CCP can be applicable.

On the most fundamental level is the structure of the nucleon. As
noted as early as in 1984~\cite{CCP}, the bag radius in the bag
model of the nucleon, naively interpreted as the confinement size,
is a gauge artifact. Here the bag boundary provides the space-time
point at which QCD and hadronic variables are matched. The CCP
statement is that the physics should not depend on where the
matching is made. How this comes about through the combination of
quantum anomalies and topology of skyrmions is detailed in
\cite{CND}. What the Cheshire-Cat phenomenon implies is that there
is a continuous map between QCD degrees of freedom and hadronic
degrees of freedom for low-energy hadronic properties.

One can think of the HLS/VM \`a la Harada and Yamawaki highlighted
in this paper as a CCP in momentum space with an additional
remarkable feature that is new, namely the presence and importance
of the vector-manifestation fixed point. The vector manifestation
(VM) may very well be present also in the chiral bag/skyrmion
picture as conjectured in \cite{LPRVchi}. The important point
regarding the VM fixed point is that while physical processes away
from the phase transition critical point -- a few of which were
considered above -- may be given by ``fusing" or ``defusing" or by
some combination of the two of the sobar and elementary-particle
modes, thereby obstructing the clear evidencing of the ``smoking
gun," the VM governs the scaling behavior of the physical
parameters that enter the processes, such as e.g. sending the
vector meson mass to zero in the chiral limit. How the chiral
symmetry restoration and nuclear interactions manifest themselves
in physical processes is therefore irrelevant.

A case that is relevant at the next level of fundamental nature is
the connection between the BR scaling $\Phi$ reflecting the
property of QCD vacuum and the Landau parameter $F_1$ as discussed
in \cite{friman,song,MR:MIGDAL}. This illustrates a possible
mapping between the complex vacuum structure of QCD characterized
by the quark condensate in medium and many-body nuclear
interactions embodied in the Landau parameters. If this
identification is correct, then it will suggest an inherent
ambiguity in delineating QCD effects from many-body hadronic
effects conventionally treated in the standard nuclear physics
approach (SNPA) discussed in Section \ref{eft}.

Although not worked out in detail, we expect the issue of
Gamow-Teller strengths in nuclei to be in a way quite similar to
the relation between the BR scaling $\Phi$ and the Landau
parameter $F_1$. It has been debated~\cite{GT} since many years as
to whether the ``quenching" of Gamow-Teller strengths in nuclei as
observed in giant Gamow-Teller resonances is due to
``conventional" multi-particle-multi-hole effects (e.g., core
polarization~\cite{arima})  or to ``exotic" effects (e.g.,
$\Delta$-hole excitations~\cite{MR73}). The resolution to this
debate~\cite{MR:telluride} is that both are relevant in a way
analogous to what happens in the Dirac phenomenology discussed in
Section \ref{diracpheno}. As noted in Section
\ref{effectiveforce}, the tensor force predominantly excites
(particle-hole) states at an energy $\sim$ 300 MeV which is
comparable to the $\Delta$-hole excitation energy. Because of
subtle cancellations between various terms of the same scale
involving multi-particle-multi-hole configurations and
$\Delta$-hole configurations, the effective Gamow-Teller coupling
constant $g_A^*$ for the transition to the lowest Gamow-Teller
state can be calculated equally well by saturating -- in the
standard EFT language -- the ``counter term" by the $\Delta$-hole
configurations or by the multi-particle-multi-hole states or by
both. Decimated down to the vicinity of the Fermi surface, both
should reside in the ``counter" term. If one were to look at the
excitation functions going over a range of energies, one of course
would have to be careful with the multitude of scales that figure
for the specific excitations involved so as not to encounter the
breakdown of the particular EFT one is using.
\subsection*{Acknowledgments}
\itt We are grateful to Ralf Rapp for providing us with his
unpublished results on the Regina and CERES processes, to Dan-Olof
Riska for useful discussions on the electromagnetic form factors
of the proton and to Chang-Hwan Lee and Byung-Yoon Park for
valuable help on the manuscript .

\end{document}